%% file: bs-bw-dw-02.tex
\documentclass[useAMS, usenatbib, a4paper]{mnras}
\usepackage{graphicx}
\usepackage{microtype}
\usepackage{xcolor}
\usepackage{fixltx2e}
\usepackage{booktabs}
\usepackage{siunitx}
\usepackage{color}
\usepackage{enumerate}
\usepackage{pdflscape}
\usepackage{rotating}
\usepackage{xr-hyper}
\usepackage{hyperref}

\usepackage[T1]{fontenc} 
\usepackage[utf8]{inputenc}

\usepackage{newtxtext}
\usepackage[varvw,smallerops]{newtxmath}

\usepackage{chemgreek}
\activatechemgreekmapping{newtx}
\usepackage{listings}

\hypersetup{colorlinks=True, linkcolor=blue!50!black, citecolor=black,
  urlcolor=blue!50!black}

\usepackage{etoolbox}
\robustify\bfseries
\robustify\itshape

\makeatletter
\patchcmd\@combinedblfloats{\box\@outputbox}{\unvbox\@outputbox}{}{%
  \errmessage{\noexpand\@combinedblfloats could not be patched}%
}%
\makeatother

\usepackage{bm}

\usepackage{aastex-compat}

\title
{Bow shocks, bow waves, and dust waves. II. Beyond the rip point}

\newcommand\AddressCRyA{Instituto de Radioastronom\'{\i}a y Astrof\'{\i}sica,
  Universidad Nacional Aut\'onoma de M\'exico, Apartado Postal 3-72,
  58090 Morelia, Michoac\'an, M\'exico}
\author[Henney \& Arthur]{
  William J. Henney\thanks{w.henney@irya.unam.mx}
  \& S. J. Arthur\\
  \AddressCRyA
}

\date{Accepted XXX. Received YYY; in original form ZZZ}

\pubyear{2019}

\input{bs-bw-dw-defs.tex}

\defcitealias{Tarango-Yong:2018a}{Paper~I}

\begin{document}
\label{firstpage}
\pagerange{\pageref{firstpage}--\pageref{lastpage}}
\maketitle
\begin{abstract}
  Dust waves are a result of gas--grain decoupling in a stream of
  dusty plasma that flows past a luminous star.  The radiation field
  is sufficiently strong to overcome the collisional coupling between
  grains and gas at a \textit{rip-point}, where the ratio of radiation
  pressure to gas pressure exceeds a critical value of roughly 1000.
  When the rip point occurs outside the hydrodynamic bow shock, a
  separate dust wave may form, decoupled from the gas shell, which can
  either be drag-confined or inertia-confined, depending on the stream
  density and relative velocity.  In the drag-confined case, there is
  a minimum stream velocity of roughly \SI{60}{km.s^{-1}} that allows
  a steady-state stagnant drift solution for the dust wave apex.  For
  lower relative velocities, the dust dynamics close to the axis
  exhibit a limit cycle behavior (rip and snap back) between two
  different radii.  Strong coupling of charged grains to the plasma's
  magnetic field can modify these effects, but for a quasi-parallel
  field orientation the results are qualitatively similar to the
  non-magnetic case. For a quasi-perpendicular field, on the other
  hand, the formation of a decoupled dust wave is strongly suppressed.
\end{abstract}

\begin{keywords}
  circumstellar matter -- radiation: dynamics -- stars: winds, outflows
\end{keywords}

\section{Introduction}
\label{sec:rip-introduction}
Stellar bow shocks are predicted to occur whenever a star moves
supersonically relative to its surrounding medium, which may either be
due to the star's own motion \citetext{runaways, e.g.,
  \citealp{Blaauw:1961a}}, or due to an independent environmental flow
\citetext{weather vanes, e.g., \citealp{Povich:2008a}}.  Bow shocks
are associated with a large variety of different types of stars, for
example: AGB stars and red super-giants \citep{Cox:2012a}; OB stars
\citep{Kobulnicky:2017a}; T~Tauri stars \citep{Gull:1979a,
  Henney:2013a}; photoevaporating protoplanetary disks
\citep{Garcia-Arredondo:2001a, Smith:2005a}; neutron stars
\citep{Cordes:1993a, Brownsberger:2014a}; planetary nebula halos
\citep{Ali:2012a}; and Galactic Center sources \citep{Geballe:2004a,
  Sanchez-Bermudez:2014a}.  The dominant emission mechanism can vary
considerably between the different classes of sources, but
recombination line radiation such as H\(\alpha\), infrared continuum
radiation from warm dust, and free-free radio continuum are common
\citep{Canto:2005a, Acreman:2016a, Meyer:2016a}.  In nearly all cases,
stellar bow shocks are easily spatially resolved and mapped, allowing
their shapes to be compared with theoretical predictions
\citep{Wilkin:1996a, Tarango-Yong:2018a}.  A different class of
stellar bow shocks are found in interacting binary systems
\citep{Stevens:1992a}, but these typically emit by non-thermal
mechanisms and are only resolvable using radio interferometry
techniques \citep{Contreras:1999a, Dzib:2013a}.

In \citet{Henney:2019a} (hereafter, Paper~I) we studied the formation
of bows around OB stars by the combined action of their winds and
radiation.  For the case of strong collisional coupling between gas
and dust grains, we showed that three different regimes of interaction
are possible, according to the relative values of stellar wind
radiative momentum efficiency \(\eta\wind\) and the UV optical depth of
the bow shell \(\tau\):
\begin{enumerate}[{\(\star\)}]
\item Wind-supported bow shocks (WBS) where \(\tau < \eta\wind\).  These are
  purely hydrodynamic (or MHD) interactions, in which the stellar
  radiation is dynamically unimportant.  This regime dominates for
  fast-moving stars and for low-density environments.
\item Radiation-supported bow waves (RBW) where
  \(\eta\wind < \tau < 1\).  In these optically thin bows, the stellar
  radiative momentum gradually decelerates the oncoming stream in a
  broad shell.  This regime is important for B~stars and weak-wind
  O~stars in moderate density environments (\(> \SI{100}{cm^{-3}}\)).
\item Radiation-supported bow shocks (RBS) where \(\tau > 1\).  These are
  optically thick shells, internally supported by the trapped stellar
  radiation pressure.  This regime applies to slow-moving O~stars in
  dense environments.
\end{enumerate}
In this paper, we investigate under what circumstances the strong
gas--grain coupling might break down, leading to a fourth regime: a
separate dust wave outside of the hydrodynamic bow shock (see Fig.~1
of Paper~I).

In neutral and molecular regions, the drag forces are relatively weak
and so divergent dynamics of gas and grains are frequently found in
the context of molecular clouds \citetext{\citealp{Hopkins:2016a,
    Lee:2017a, Mattsson:2019a}, but see \citealp{Tricco:2017a}} and
protoplanetary disks \citep{Weidenschilling:1977b, Birnstiel:2010a,
  Dipierro:2018a}.  In ionized regions, the electrostatic forces
between charged dust grains and charged gas particles (principally
protons) leads to greatly increased drag forces, which tend to
maintain a tight coupling between dust and gas \citep{Draine:2011a}.
However, in a photoionized \hii{} region around a high-mass star, the
dust feels a much larger radiation force than does the gas. This leads
to a slow relative drift between the two components in the outer
regions of the nebula \citep{Gail:1979a, Akimkin:2015a, Akimkin:2017a,
  Ishiki:2018a} and a total decoupling very close to the central star
\citep[Fig.~8 of][]{Draine:2011a}.

With respect to stellar bow shocks, gas--grain decoupling has been
previously studied in the context of post-shock flow, where it is the
sudden deceleration of the gas that is the primary impetus for the
dust to separate.  The case of a cool red supergiant was simulated by
\citet{van-Marle:2011a}, where large grains present in the stellar
wind are not stopped in the inner termination shock, but are carried
through the contact discontinuity by their inertia, penetrating into
the interstellar medium.  \citet{Katushkina:2017a} simulated the
opposite case, where grains in the ambient medium become decoupled
from the gas after passing the outer bow shock, and subsequently
gyrate about the magnetic field, forming filamentary structures (see
also \citealp{Katushkina:2018a}).  In this paper, we study a different
decoupling mechanism, where it is the stellar radiation force acting
on the grains that causes a separation from the gas \emph{before the
  stream reaches the bow shock}.  This mechanism was first proposed by
\citet{Ochsendorf:2014b} to explain the infrared emission arc around
the high-mass multiple system \(\sigma\)~Ori.

The plan of the paper is as follows.
In \S~\ref{sec:recap-paper-i} we give some results from Paper~I that
we will need in later sections.
In \S~\ref{sec:gas-free-bow} we calculate analytical models of the
shape of dust waves in the drag-free limit.
%
In \S~\ref{sec:imperf-coupl-betw} we calculate in detail the grain
charging and dynamics, subject to radiation and drag forces, in order
to determine the \textit{rip point}, which is where gas--grain coupling
catastrophically breaks down.  This is used to determine existence
conditions for the presence of a decoupled dust wave.
In \S~\ref{sec:magn-effects-grain} we consider the coupling between
the grains and the plasma's magnetic field, and how this effects the
existence and structure of dust waves.
In \S~\ref{sec:discussion} we discuss our results in the context of
previous studies.
In \S~\ref{sec:summary} we summarise our findings.
In Appendix~\ref{sec:equat-moti-grains} we provide additional
technical information on the numerical calculation of grain
trajectories.

\section{Recapitulation of Paper~I}
\label{sec:recap-paper-i}

In Paper~I, we found approximate expressions for the bow radius in each of the three regimes discussed in the introduction: 
\begin{equation}
\begin{aligned}
  \text{RBS} \quad& (\tau_*^2 > 1): & R_0 &\approx  (1 + \eta\wind)^{1/2} \, R_*  \\
  \text{RBW} \quad& (\eta\wind < \tau_*^2 < 1): & R_0 &\approx 2 \, \tau_* \, R_* \\
  \text{WBS} \quad& (\tau_*^2 < \eta): & R_0 &\approx \eta\wind^{1/2} \, R_*  
\end{aligned}\label{eq:x-cases}
\end{equation}
In these expressions, we use a fiducial radius
\begin{equation}
  \label{eq:Rstar}
  R_* = \left(\frac{L}{4\pi c \rho v_\infty^2}\right)^{1/2} \ ,
\end{equation}
a fiducial optical depth
\begin{equation}
  \label{eq:tau-star}
  \tau_* = \rho \kappa R_* \ ,
\end{equation}
and a factor that describes the efficiency of the stellar wind as the fraction of the stellar radiation momentum that is converted to wind momentum
\begin{equation}
  \label{eq:wind-efficiency}
  \eta\wind  = \frac{c\, \dot{M} V\wind}{L} \ .
\end{equation}
Paper~I's equation~(11) allows a more exact value for \(R_0\) to be
found numerically in intermediate cases.  Note that \(R_0\) is the
star--apex distance measured along the symmetry axis of the bow
\citetext{see \citealp{Tarango-Yong:2018a} for explanation of
  nomenclature and discussion of bow shapes}, and is calculated in the
limit that the momentum transfer occurs at a surface.  In cases where
the shell's finite width is significant, \(R_0\), should correspond
approximately to the astropause (contact discontinuity) in the WBS
regime, or a UV optical depth of unity (as measured from the star) in
the RBS regime.  The total perpendicular optical depth of the shell to
stellar radiation can be found as
\begin{equation}
  \label{eq:actual-tau}
  \tau = \frac{R_0}{R_*} \, \tau_* \ .
\end{equation}

In the foregoing, we employ the stellar bolometric luminosity, \(L\),
wind mass-loss rate, \(\dot{M}\), and terminal velocity, \(V\wind\),
together with the ambient stream's mass density, \(\rho\), relative
velocity \(v_\infty\), and effective dust opacity, \(\kappa\).  It is convenient
to define dimensionless versions of these parameters by normalizing to
typical values:
\begin{align*}
  \dot{M}_{-7} &= \dot{M} / \bigl(\SI{e-7}{M_\odot.yr^{-1}}\bigr) \\
  V_3 &= V / \bigl(\SI{1000}{km.s^{-1}}\bigr) \\
  L_4 &= L / \bigl(\SI{e4}{L_\odot}\bigr) \\
  v_{10} &= v_\infty / \bigl( \SI{10}{km.s^{-1}} \bigr) \\
  n &= (\rho / \bar{m}) / \bigl( \SI{1}{cm^{-3}} \bigr) \\
  \kappa_{600} &= \kappa / \bigl( \SI{600}{cm^2.g^{-1}} \bigr) \ ,
\end{align*}
where \(\bar{m}\) is the mean mass per hydrogen nucleon
(\(\bar{m} \approx 1.3 m_{\text{p}} \approx \SI{2.17e-24}{g}\) for solar
abundances).  In terms of these dimensionless parameters,
equations~(\ref{eq:Rstar}--\ref{eq:wind-efficiency}) take the
convenient forms:
\begin{align}
  \label{eq:Rstar-typical}
  R_* / \si{pc} &= \num{2.21} \, (L_4 / n)^{1/2} \,v_{10}^{-1} \\
  \label{eq:taustar-typical}
  \tau_* &= \num{0.0089} \,\kappa_{600} \, (L_4 \,n)^{1/2} \,v_{10}^{-1} \\
  \label{eq:wind-eta-typical}
  \eta\wind &= \num{0.495} \,\dot{M}_{-7} \,V_3  \,L_4^{-1} \ .
\end{align}

\input{app-dust-wave}

\input{sec-weak-coupling}

\input{sec-dust-wave-discussion}

\section*{Acknowledgements}
We are grateful for financial support provided by Dirección General de
Asuntos del Personal Académico, Universidad Nacional Autónoma de
México, through grants Programa de Apoyo a Proyectos de Investigación
e Inovación Tecnológica IN112816 and IN107019.  This work has made
extensive use of Python language libraries from the SciPy
\citep{Jones:2001a} and AstroPy \citep{Astropy-Collaboration:2013a,
  Astropy-Collaboration:2018a} projects.  We thank Olga Katushkina for useful discussions.

\bibliographystyle{mnras}
\bibliography{bowshocks-biblio}
\appendix
\input{app-dust-equations}
\bsp	
\label{lastpage}
\end{document}

%% file: bs-bw-dw-defs.tex
\DeclareMathOperator\erf{erf}

\providecommand{\abs}[1]{\lvert#1\rvert}

\newcommand\Qp{\ensuremath{Q_{\text{p}}}}
\newcommand\Qpbar{\ensuremath{\bar{Q}_{\text{p}}}}
\newcommand{\grain}{\ensuremath{_{\text{d}}}}
\newcommand{\B}{\ensuremath{_{\scriptscriptstyle\text{B}}}}
\newcommand{\alfven}{\ensuremath{_{\scriptscriptstyle\text{A}}}}

\newcommand\frad{\ensuremath{f_{\text{rad}}}}

\newcommand\thm{\ensuremath{\theta_{\text{m}}}}
\newcommand\drag{\ensuremath{_{\text{drag}}}}
\newcommand{\gas}{\ensuremath{_{\text{gas}}}}
\newcommand{\wind}{\ensuremath{_{\text{w}}}}

\newcommand{\drift}{\ensuremath{_{\text{drift}}}}
\newcommand\rad{\ensuremath{_{\text{rad}}}}

\newcommand\Rmin{\ensuremath{R_{\scriptscriptstyle\text{min}}}}
\newcommand\sound{\ensuremath{c_{\text{s}}}}

\newcommand\starstar{\ensuremath{_{**}}}

\newcommand\hii{\ion{H}{ii}}

\newcommand\amu{\ensuremath{a_{\si{\um}}}}

\newcommand\freeze{\ensuremath{_{\text{tight}}}}


%% file: app-dust-wave.tex
\section{Dust waves in the drag-free limit}
\label{sec:gas-free-bow}


\begin{figure}
  (a)\\
  \includegraphics[width=\linewidth]{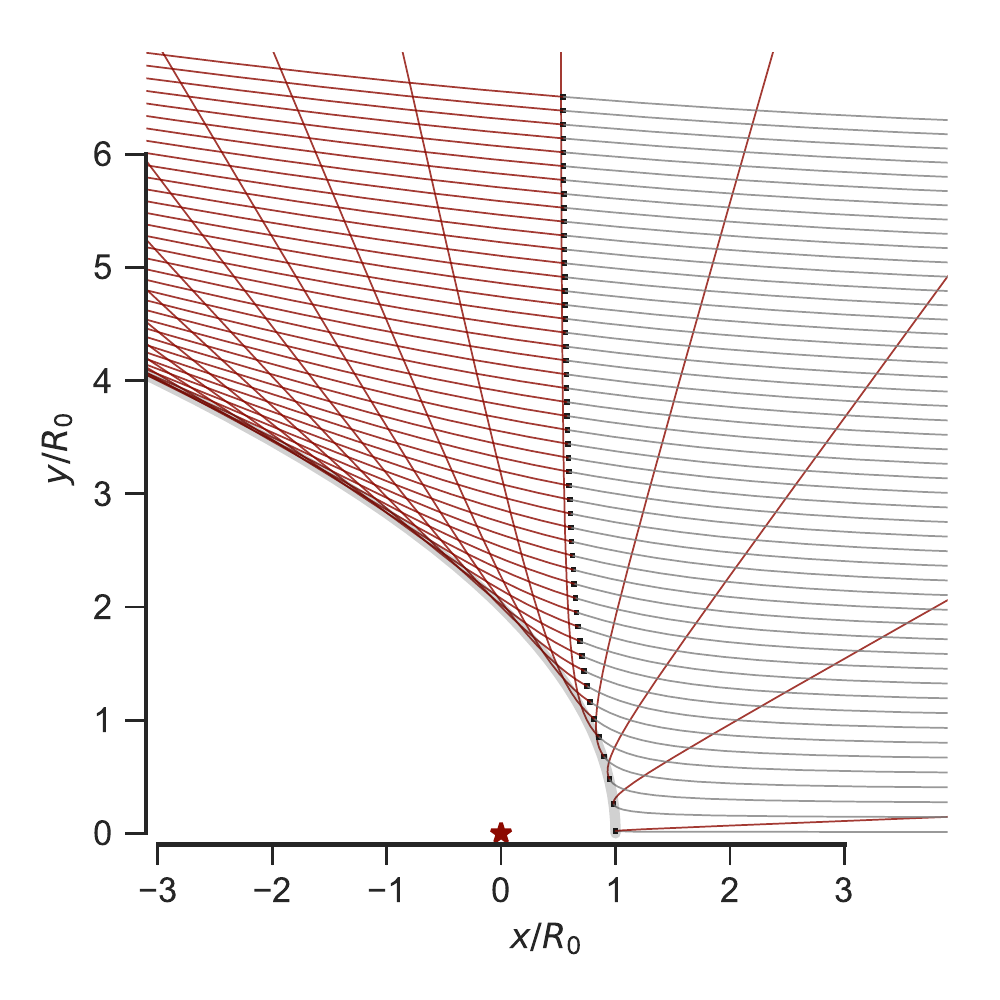}
  (b)\\
  \includegraphics[width=\linewidth]{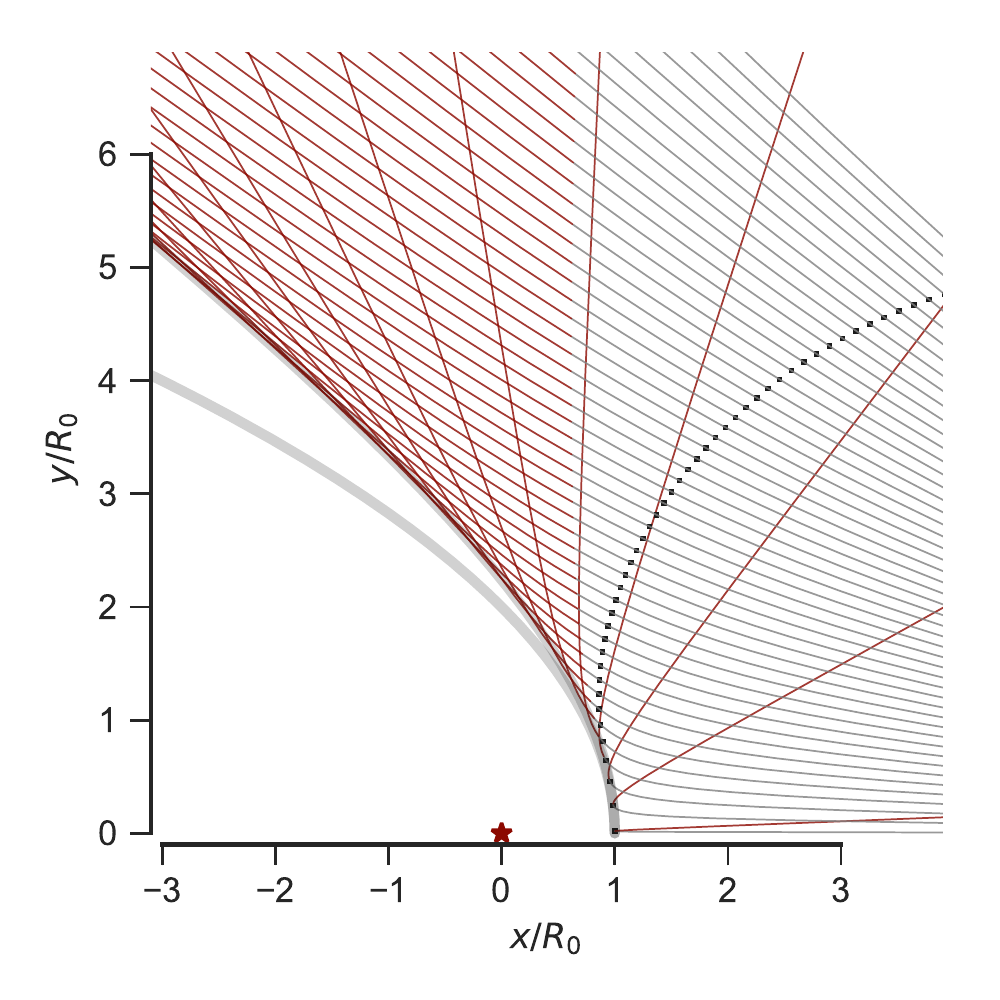}
  \caption[Dust grain trajectories]{Dust grain trajectories under
    influence of a repulsive central \(r^{-2}\) radiative force.
    (a)~A parallel stream of dust grains approach from the right at a
    uniform velocity and with a variety of impact parameters (initial
    \(y\)-coordinate). The central source is marked by a red star at
    the origin, and its radiative force deflects the trajectories into
    a hyperbolic shape, each of which reaches a minimum radius marked
    by a small black square.  The incoming hyperbolic trajectories are
    traced in gray and the outgoing trajectories are traced in red.
    The locus of closest approach of the outgoing trajectories is
    parabolic in shape (traced by the thick, light gray line) and this
    constitutes the inner edge of the bow wave.  (b)~The same but for
    a divergent stream of dust grains that originates from a source on
    the \(x\) axis at a distance \(D = 10 R\starstar\) from the
    origin.  In this case, the inner edge of the bow wave is
    hyperbolic and the parallel stream result is also shown (inner
    thick light gray line) for comparison.}
  \label{fig:dust-trajectories}
\end{figure}

In the optically thin limit, a spherical dust grain of radius \(a\)
situated a distance \(R\) from a point source of radiation with
luminosity spectrum \(L_\nu\) will experience a repulsive, radially
directed radiative force \citep[e.g.,][]{Spitzer:1978a}
\begin{equation}
  \label{eq:dust-rad-force}
  \frad = \frac{\pi a^2 } {4 \pi R^2 c}  \int_0^\infty \Qp \, L_\nu \, d\nu 
\end{equation}
where \Qp{} is the frequency-dependent radiation pressure efficiency\footnote{%
  \label{fn:Qp}
  For absorption efficiency \(Q_{\text{abs}}\), scattering efficiency
  \(Q_{\text{scat}}\), and asymmetry parameter (mean scattering
  cosine) \(g\), we have
  \(\Qp = Q_{\text{abs}} + (1 - g) Q_{\text{scat}}\) \citep[e.g.,
  \S~4.5 of][]{Bohren:1983a}.} %
of the grain, and \(c\) is the speed of light.


If \(\frad\) is the only force experienced by the grain, then it will
move on a \textit{ballistic} trajectory, determined by its initial
speed at large distance, \(v_\infty\), and its impact parameter, \(b\).
For \(b = 0\), the grain radially approaches the source with initial
radial velocity \(-v_\infty\), which is decelerated to zero at the distance
of closest approach, \(R\starstar\), given by energy conservation:
\begin{equation}
  \label{eq:dust-r0}
  R\starstar = \frac{\kappa\grain L} {2 \pi c v_\infty^2} \ ,
\end{equation}
where we have defined a frequency-averaged single-grain opacity
(\si{cm^2.g^{-1}}) as
\begin{equation}
  \label{eq:kappa-grain}
  \kappa\grain = \frac{\pi a^2}{m\grain L} \int_0^\infty \Qp \, L_\nu \, d\nu \ ,
\end{equation}
in which \(m\grain\) is the grain mass and \(L\) is the bolometric
source luminosity.  The grain then turns round and recedes from the
source along the same radius, reaching a velocity of \(+v_\infty\) at large
distance.  Note that \(R\starstar\) as given by
equation~\eqref{eq:dust-r0} is almost the same as Paper~I's
equation~(6), but with the important difference that it is for a
single grain considered in isolation, rather than a well-coupled dusty
plasma.  Since we are here ignoring collisional effects, there is no
pressure, and therefore nothing to stop the spatial coexistence of an
inbound and outbound dust stream.  For the well-coupled case, this is
not possible and a shocked shell must form (see Paper~I's Fig.~5b).

For \(b > 0\), the problem is formally identical to that of Rutherford
scattering, or (modulo a change of sign) planetary orbits.  The method
of solution (via introduction of a centrifugal potential term and
reduction to a 1-dimensional problem) can be found in any classical
mechanics text \citep[e.g.,][\S~14]{Landau:1976a}.  The trajectory,
\(R\grain(\theta; b)\), is found to be hyperbolic, characterized by an
eccentricity,
\(\varepsilon = \bigl( 1 + 4 b^2 / R\starstar^2\bigr)^{1/2}\), and polar angle
of closest approach, \(\thm = \cos^{-1} \varepsilon^{-1}\).  The trajectory is
symmetrical about \(\thm\) and can be written as
\begin{equation}
  \label{eq:dust-r-theta}
  \frac{R\grain(\theta; b)} {R\starstar} = 
  \frac{ \tfrac12 \bigl( \varepsilon^2 - 1 \bigr)} {\varepsilon \cos(\theta - \thm) - 1} \ , 
\end{equation}
with a total deflection angle of \(\ang{180} - 2 \thm\), which is equal to
\ang{90} when \(b = 0.5 R\starstar\).

\subsection{Parallel dust stream}
\label{sec:dust-parallel}

If the incoming dust grains initially travel along parallel
trajectories with varying \(b\) but the same \(v_\infty\), then deflection
by the radiative force will form a bow-shaped dust wave around the
radiation source, as shown in Figure~\ref{fig:dust-trajectories}.
However, the inner edge of the dust wave, \(R_{\text{in}}(\theta)\) is not
given by the closest approach along individual trajectories,
\(R\grain(\thm; b)\), but instead must found by minimizing
\(R\grain(\theta; b)\) over all \(b\) for each value of \(\theta\), which yields
\begin{equation}
  \label{eq:dust-r-in}
  \frac{R_{\text{in}}(\theta)} {R\starstar} = \frac{2}{1 + \cos\theta} \ .
\end{equation}
This is the polar form of the equation for the confocal parabola,
which is discussed in detail in \S~\ref{Q-sec:conic} and
Appendix~\ref{Q-app:parabola} of \citet{Tarango-Yong:2018a}.  In that
paper, the dimensionless quantities \textit{planitude} \(\Pi\) and
\textit{alatude} \(\Lambda\) are introduced as a way of efficiently
characterizing bow shapes.  The confocal parabola has planitude and
alatude of \(\Pi = \Lambda = 2\) and these are unchanged under projection at
any inclination.

\subsection{Divergent dust stream}
\label{sec:dust-divergent}

If the dust grains are assumed to originate from a second point
source, located at a distance \(D\) from the radiation source, then
the incoming stream will be divergent instead of plane parallel.  The
individual streamlines are not affected by this change and are still
described by equation~\eqref{eq:dust-r-theta}, except that the
trajectory axes for \(b > 0\) are no longer aligned with the global
symmetry axis, so we must make the substitution
\(\theta \to \theta + \theta_1(b)\), where \(\sin \theta_1 = b / D\) (see
Fig.~\ref{Q-fig:crw-schema} of \citealp{Tarango-Yong:2018a}). We
parameterize the degree of divergence as \(\mu = R\starstar / D\) and, as
before, \(R\grain(\theta + \theta_1(b, \mu); b)\) is minimized over all
trajectories to find the shape of the bow wave's inner edge.  This
time, the result is a confocal hyperbola:
\begin{equation}
  \label{eq:dust-divergent-r-in}
  \frac{R_{\text{in}}(\theta; \mu)} {R\starstar} = \frac{1 + \varepsilon_\mu}{1 + \varepsilon_\mu\cos\theta} \ ,
\end{equation}
where the eccentricity is (to first order in \(\mu\))
\( \varepsilon_\mu = (1 - 2\mu)^{-1}\).  An example is shown in
Figure~\ref{fig:dust-trajectories}b for \(\mu = 0.1\).  Unsurprisingly,
the resulting bow shape is more open than in the parallel stream case,
increasingly so with increasing \(\mu\).  The planitude and alatude are
both equal: \(\Pi = \Lambda = 1 + \varepsilon_\mu\)
(\(\Pi = \Lambda = 2.25\) in Fig.~\ref{fig:dust-trajectories}b).

\subsection{Magnetized dust waves}
\label{sec:tight-magn-coupl}

\begin{figure}
  \centering
  \includegraphics[width=\linewidth]{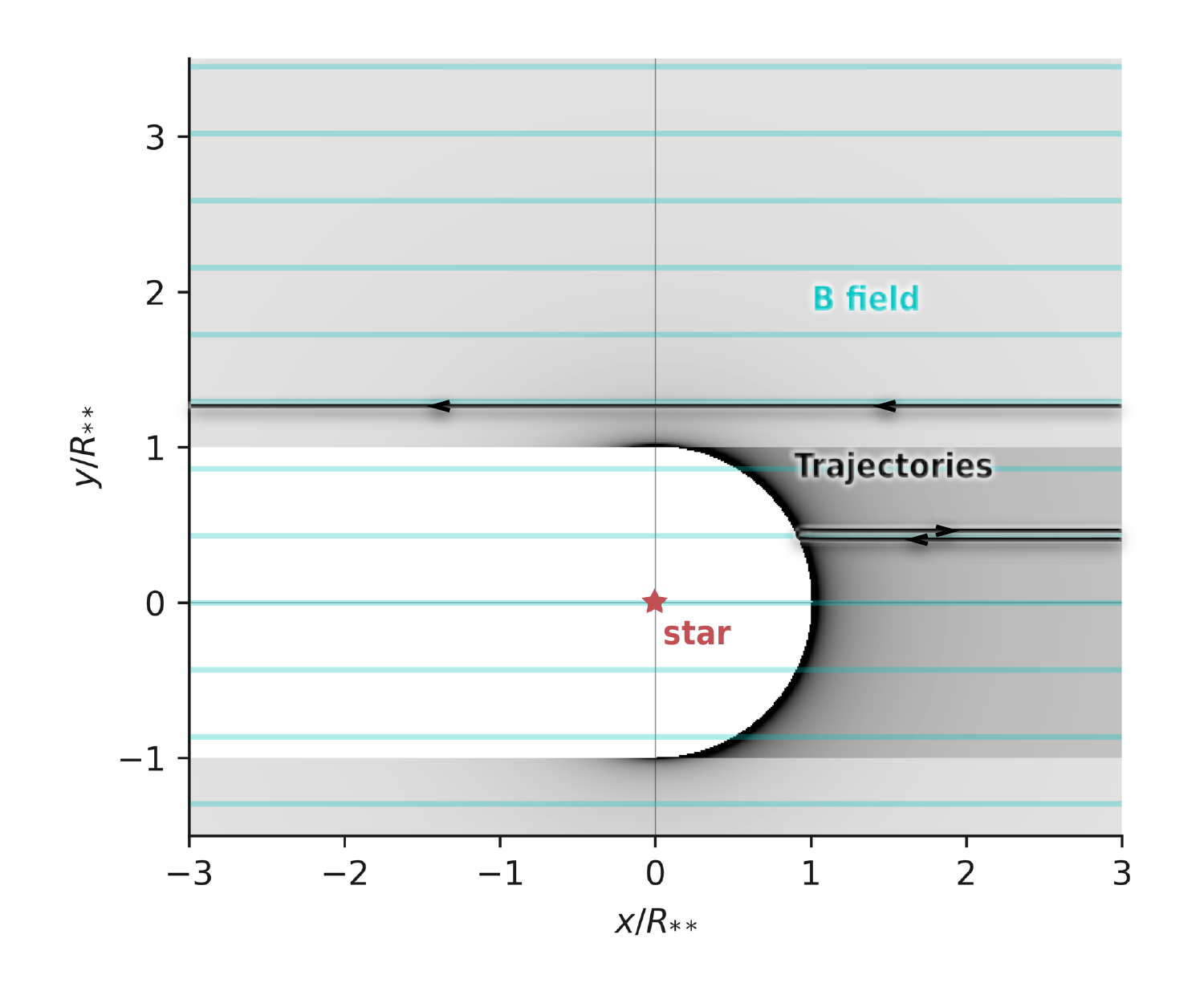}
  \caption{Dust wave formed by action of radiation forces on grains
    that are tightly coupled to a uniform parallel magnetic field.
    Two example trajectories, one with \(b > R\starstar\) (upper) and
    one with \(b < R\starstar\) (lower) are shown schematically in
    orange.  The orientation of the magnetic field lines is shown in
    blue.  The grayscale image shows the resultant dust density
    distribution.}
  \label{fig:inertia-thB0}
\end{figure}

\begin{figure}
  \centering
  \includegraphics[width=\linewidth]{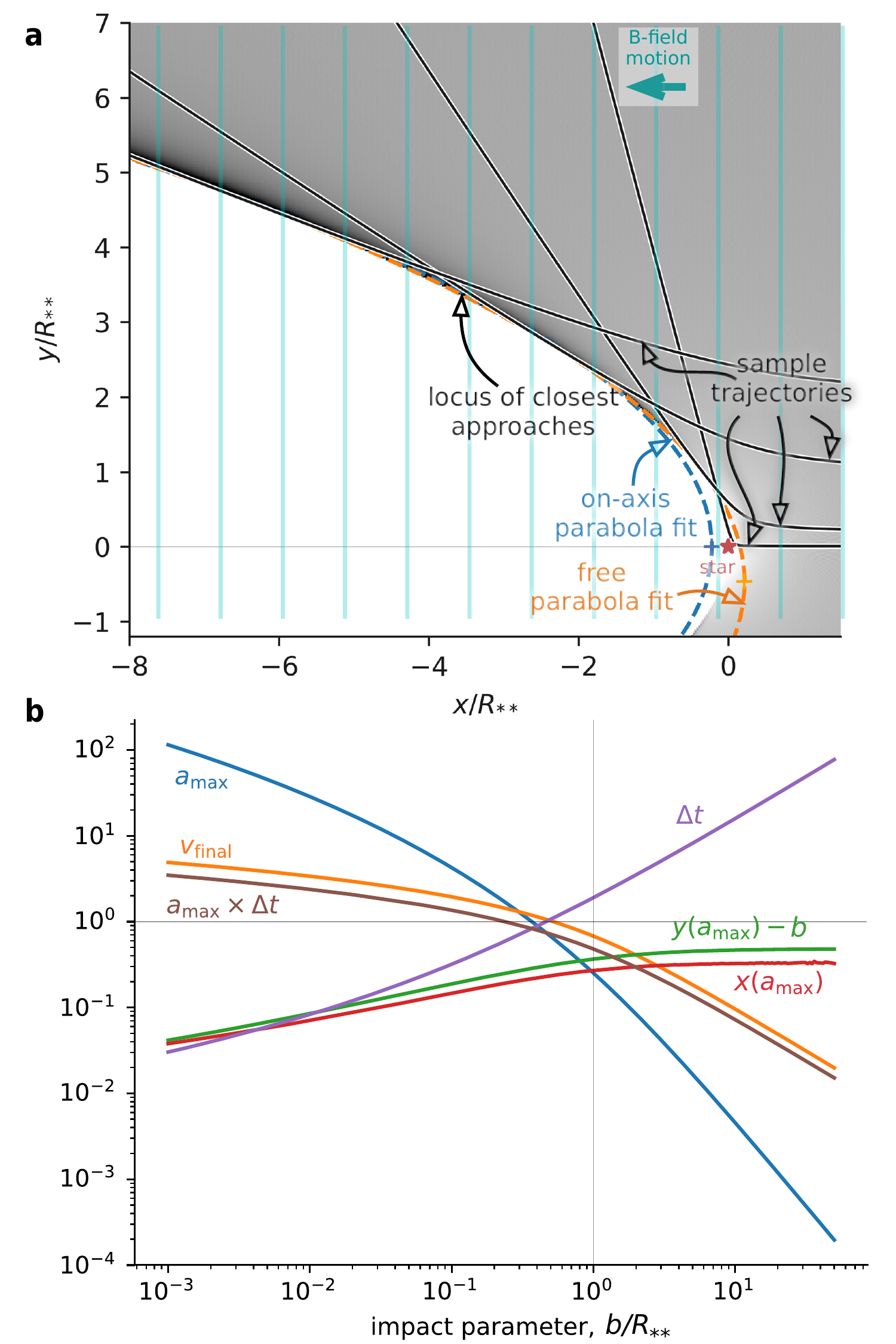}
  \caption{Dust wave formed by action of radiation forces on grains
    that are tightly coupled to a uniform perpendicular magnetic
    field.  (a)~Sample trajectories are shown by thin black lines and
    the resultant dust density distribution in grayscale.  Two
    parabolic fits to the inner edge of the dust wave in the wing
    region (\(y > R\starstar\)) are shown. The orange line shows a
    simultaneous fit to both wings, while the green line shows a fit
    to a single wing, in which the parabola apex is not constrained to
    lie on the axis.  The second fit has much smaller residuals,
    indicating that the overall dust wave shape is ``pointier'' than a
    parabola, but this is hard to quantify because the disappearance
    of the dense shell in the apex region makes it impossible to
    define \(R_0\) for the bow.  (b)~Trajectory parameters over a wide
    range of impact parameters, shown on a log--log scale. Distances
    are in units of \(R\starstar\), times are in units of
    \(R\starstar / v_\infty\), velocities are in units of \(v_\infty\), and
    accelerations are in units of \(v_\infty^2 / R\starstar\). The grain's
    acceleration along the \(y\) axis has a maximum value,
    \(a_{\text{max}}\), which occurs at a position
    \(x(a_{\text{max}}), y(a_{\text{max}})\), just before the grain is
    swept past the star, with a duration (FWHM) of \(\Delta t\).  The
    grain's asymptotic \(y\) velocity is \(v_{\text{final}}\) and the
    fact that this closely tracks \(a_{\text{max}} \times \Delta t\) indicates
    that the majority of the acceleration occurs in a sharp impulse.
    The curves tend to straight lines at the right side of the graph,
    which gives the asymptotic scaling relations discussed in the
    text.}
  \label{fig:inertia-thB90}
\end{figure}

We will show in \S~\ref{sec:magn-effects-grain} that for
sufficiently small grains the Larmor magnetic gyration radius,
\(r\B\), is small compared with other length scales of interest, and
so the grains are effectively tied to the magnetic field lines.  In
this approximation, we can calculate the grain dynamics in the
drag-free limit using \(f\rad\) as the sole force as above, but this
time allowing acceleration only along the field lines.  Assuming a
uniform magnetic field, the only extra parameter needed is
\(\theta\B\), the angle between the field direction and the direction of
the dust stream (assumed to be plane parallel).

We now calculate in detail the grain trajectories in this limit for
two cases, with the magnetic field oriented parallel and perpendicular
to the stream direction, respectively. In both cases, the stream
trajectories are assumed parallel to one another, as in
\S~\ref{sec:dust-parallel}. These two cases are sufficient to give a
flavor of the effects of a magnetic field on the dust wave structure.
Further models at intermediate angles, and which also include the
effects of gas drag, are presented in \S~\ref{sec:magn-effects-grain}.

\subsubsection{Parallel magnetic field}
\label{sec:parall-magn-field}

For \(\theta\B = 0\), the \(b = 0\) trajectory is identical to the
non-magnetic case since the grain velocity and radiation force are
both parallel to the field, which yields zero Lorentz force and zero
\(\bm{f} \times \bm{B}\) drift. Therefore, the axial turnaround radius,
\(R\starstar\), is still given by equation~\eqref{eq:dust-r0},
whatever the value of \(r\B\).  For \(b > 0 \), \(f\rad\) has
a component perpendicular to \(\bm{B}\), which will induce a helical
gyromotion around the field lines, but the guiding center must move
parallel to the axis if \(r\B \ll R\starstar\).  Thus, the guiding
center motion can be found from conservation of potential plus kinetic
energy in one dimension:
\begin{equation}
  \label{eq:parallel-energy-balance}
  \frac{v^2}{v_\infty^2} + V\rad = 1 \ , 
\end{equation}
where \(V\rad\) is a suitably normalized potential of the projected
radiation force along the field lines (\(x\) axis, where
\(x = R \cos\theta = b \cot\theta\)):
\begin{equation}
  \label{eq:parallel-radiation-potential}
  V\rad = R\starstar \int_x^\infty \frac{\cos\theta}{R^2}\, dx
  = \frac{R\starstar\sin\theta}{b}
  = \frac{R\starstar}{R} \ .
\end{equation}
From equation~\eqref{eq:parallel-energy-balance}, the trajectory must
turn around when \(V\rad = 1\), and
equation~\eqref{eq:parallel-radiation-potential} shows that this
occurs at the same spherical radius, \(R = R\starstar\), for all
impact parameters, \(b\), so that the inner boundary of the dust wave
is hemispherical in shape:
\begin{equation}
  \label{eq:thB-0-shape}
  \frac{R_{\text{in}}(\theta)} {R\starstar} = 1 \ ,
\end{equation}
yielding planitude and alatude of \(\Pi = \Lambda = 1\).  Note, however, that
this only applies to streamlines with \(b \le R\starstar\).  For those
with \(b > R\starstar\), the maximum \(V\rad\), which occurs at
\(x = 0\), is smaller than unity, so that grains on these streamlines
do not turn around, although they do slow down temporarily as they go
past the star.

The grain density of the inflowing stream follows from mass continuity as:
\begin{equation}
  \label{eq:thB-0-density}
  n\grain(R) = \frac{n \bar{m} Z\grain}{m\grain}
  \left( 1 - \frac{R\starstar}{R} \right)^{-1/2} \ ,
\end{equation}
where for simplicity we assume a single grain species of mass
\(m\grain\) and dust-gas ratio \(Z\grain\). The outflowing stream has
exactly the same velocity profile as the inflowing one, apart from a
change of sign for those streamlines that turn round.  Therefore, in
the region where the two streams co-exist (\(y \le R\starstar\),
\(x > 0\), and \(R > R\starstar\)), the total density is double that
given by equation~\eqref{eq:thB-0-density}.  This is illustrated in
Figure~\ref{fig:inertia-thB0}.

\subsubsection{Perpendicular magnetic field}
\label{sec:perp-magn-field}

For \(\theta\B = \ang{90}\), the guiding center is forced to move at a
constant speed in the \(x\) direction, so that \(x = -v_\infty t\), while
the motion in the \(y\) direction obeys the ODE:
\begin{equation}
  \label{eq:ode-perp-bfield}
  \frac{d^2 y}{d x^2} = \tfrac12 R\starstar  \,
  y \, \bigl( x^2 + y^2\bigr)^{-3/2} \ .
\end{equation}
We have been unable to find an analytic solution to this equation, but
a numerical solution is shown in Figure~\ref{fig:inertia-thB90}a.
When the impact parameter is larger than \(b \sim R\starstar\), the
trajectories are very similar to in the non-magnetic case
(Fig.~\ref{fig:dust-trajectories}a).  As shown in
Figure~\ref{fig:inertia-thB90}b, the interaction of the grain with the
radiation field in this large-\(b\) regime can be approximated as an
impulsive acceleration of magnitude \(\sim b^{-2}\) and duration
\(\sim b\), producing a final \(y\) velocity of \(\sim b^{-1}\).  Since the
\(x\) velocity is constant, the total deflection angle is also of
order \(\sim b^{-1}\).  The overlap of the outgoing trajectories produces
a dense concentration of grains at the inner edge, which is roughly
parabolic in shape.  However, for \(b < R\starstar\) the remorseless
advance of the magnetic field does not allow the grains to slow down
and turn round, as they do in the non-magnetic and parallel-field
cases.  As a result, no dense shell forms in the apex region, but
instead there is a diffuse minimum in the density of grains around the
star due to the high grain velocities reached there.  This means that
the apparent morphology of a pure dust wave becomes very sensitive to
the radial dependence of the grain emissivity.  If this is
sufficiently steep, then the apex would coincide with the position of
the star, although in practice the presence of a wind-supported bow
shock, however small, will complicate the picture.  Since there is no
unambiguous definition of the star--apex distance in this case, it is
not possible to define the shape parameters \(\Pi\) and \(\Lambda\) either.


%% file: sec-weak-coupling.tex
\section{Gas--grain coupling and decoupling}
\label{sec:imperf-coupl-betw}

\begin{figure}
  \includegraphics[width=\linewidth]{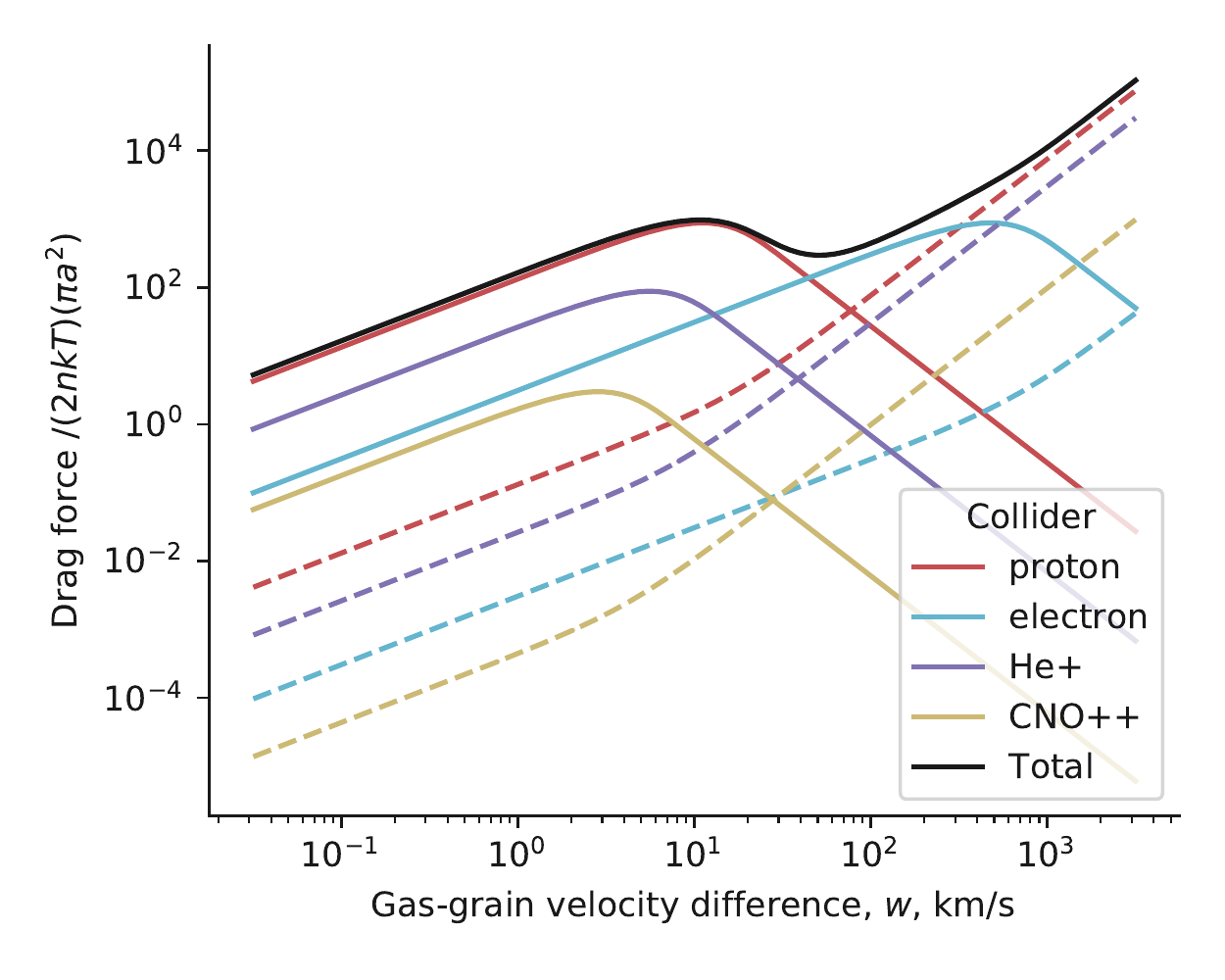}
  \caption{Contributions of different collider species to the
    dimensionless drag force, \(f\drag / f_*\), as a function of
    gas--grain slip velocity, \(w\).  Solid lines show the Coulomb
    (electrostatic) drag, while dashed lines show the Epstein
    (solid-body) drag.  Results are shown for dimensionless grain
    potential \(\phi = 10\).  All Coulomb forces scale with
    \(\phi^2\), while the Epstein forces are independent of \(\phi\).  The
    species labelled ``CNO++'' represents the combined effect of all
    metals (see App.~\ref{sec:equat-moti-grains}).}
  \label{fig:drag-components}
\end{figure}

\begin{figure}
  \includegraphics[width=\linewidth]{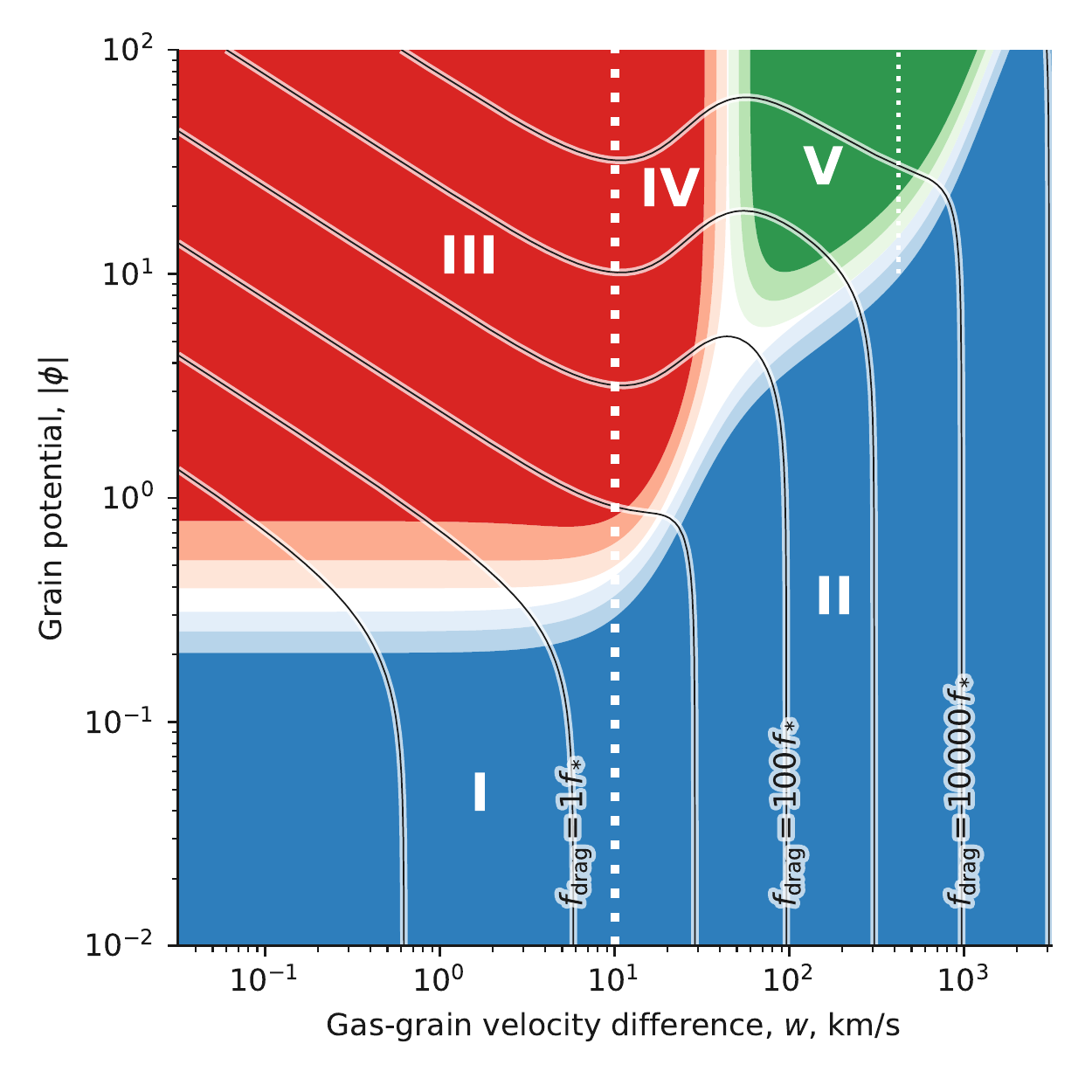}
  \caption{Regimes of gas--grain drag as a function of slip velocity
    and grain potential.  The different regimes are indicated by bold
    roman numerals, as explained in
    Table~\ref{tab:fdrag-regimes}. Blue shading indicates regions
    dominated by Epstein (solid-body) drag, whereas red and green
    shading indicate regions dominated by Coulomb drag due to protons
    and electrons, respectively.  In each case, the saturated color
    represents a contribution \(> 70\%\) of the relevant component to
    the total drag force, while progressively lighter shading
    represents the \(> 60\%\) and \(> 50\%\) levels.  The thick white
    dotted line indicates the transition between the subthermal and
    superthermal regimes for protons, while the thin white dotted line
    indicates the corresponding transition for electrons.  Contours
    show the total drag force in units of \(f_*\) (see
    eq.~[\ref{eq:fstar}]) in decade intervals from \(0.1\) to
    \(10^4\), as labelled.  Results are shown for
    \(T = \SI{8000}{K}\) and \(n = \SI{100}{cm^{-3}}\), but the
    differences are very slight throughout the ranges
    \(T = \text{\SIrange{5000}{15000}{K}}\) and
    \(n = \text{\SIrange{e-3}{e6}{cm^{-3}}}\).}
  \label{fig:drag-v-phi-plane}
\end{figure}


We now consider how the results of the previous section are changed
when drag forces from the gas are taken into account.  After
considering the general form of these forces in
\S~\ref{sec:drag-force-grains}, we study the behavior of the grain
electrostatic potential in \S~\ref{sec:cloudy-models-dust} and the
nature of the decoupling that occurs for sufficiently strong radiation
fields in \S~\ref{sec:gas-grain-separ}.  In
\S~\ref{sec:exist-cond-separ} we use this information to deduce
conditions for the existence of dust waves, and calculate grain
trajectories in \S~\ref{sec:two-regimes-post} and estimate the
back-reaction on the plasma in \S~\ref{sec:back-reaction-gas}.


\subsection{Drag force on grains}
\label{sec:drag-force-grains}

\begin{table}
  \centering
  \caption{Regimes of drag force as function of grain potential and slip speed}
  \label{tab:fdrag-regimes}
  \renewcommand\arraystretch{1.3}
  \resizebox{\linewidth}{!}{%
    \begin{tabular}{@{}r l l l@{}}
    \toprule
      & Regime & Approximate criteria & \(f\drag / f_*\) \\ \midrule
      I & Epstein subsonic & \(\phi^2 \ll 1\)
                             and \(w_{10} < 1\) & \(1.5\, w_{10}\) \\
      II & Epstein supersonic & \(w_{10} > 1\)
                                and \(w_{10} > 5\,\abs{\phi}\)& \( w_{10}^2\) \\
      III & Coulomb p\(^+\) subthermal & \(\phi^2 > 1\)
                                         and \(w_{10} < 1\) & \((1 + 20\, \phi^2)\,
                                                              w_{10}\) \\
      IV & Coulomb p\(^+\) superthermal & \(\phi^2 > 1\)
                                          and \(1 < w_{10} < 5\) & \(w_{10}^2
                                                                   + 10\, \phi^2/w_{10}^2 \) \\
      V & Coulomb e\(^-\) subthermal & \(\phi^2 > 20\)
                                 and \(5 < w_{10} < 42\) & \(0.48\, \phi^2 \,
                                                           w_{10}\) \\
    \bottomrule
  \end{tabular}
  }
\end{table}

The drag force \(f\drag\) on a charged dust grain moving at a relative
speed \(w\) through a plasma has contributions from both direct
collisions and from electrostatic Coulomb interactions with ions and
electrons \citep{Draine:1979a}.  Full details of the equations and
collider species used are given in
Appendix~\ref{sec:equat-moti-grains}. Results are shown in
Figure~\ref{fig:drag-components}, where dashed lines correspond to
direct solid body collisions and solid lines to electrostatic
interactions.  The latter depend on the grain potential, which is
described in dimensionless terms by \(\phi\), the electrostatic potential
energy of a unit charge at the surface of a grain of charge
\(z\grain\) and radius \(a\), in units of the characteristic thermal
energy of a gas particle:
\begin{equation}
  \label{eq:phi-potential}
  \phi = \frac{e^2 z\grain}{a kT} \ .
\end{equation}
The electrostatic contributions to \(f\drag\) are proportional to
\(\phi^2\) (results are shown for \(\abs{\phi} = 10\)), whereas the
solid-body contributions are independent of \(\phi\).  The drag force is
put in dimensionless units by dividing by a characteristic force:
\begin{equation}
  \label{eq:fstar}
  f_* = 2 n k T \cdot \pi a^2 \ , 
\end{equation}
which is approximately the ionized gas pressure multiplied by the
grain geometric cross section.

For grains with low electric charge, \(\phi^2 \ll 1\), the drag force is
dominated by direct collisions of protons with the grain (dashed red
line in Fig.~\ref{fig:drag-components}).  The gas collisional mean
free path is much larger than the grain size, so the drag is in the
Epstein regime \citep{Weidenschilling:1977b}.
As the relative gas--grain slip speed, \(w\), increases, \(f\drag\)
first increases linearly with \(w\) reaching \(f\drag \approx f_*\) at
\(w = \sound \approx \SI{10}{km.s^{-1}}\), then transitions to a quadratic
increase in the supersonic regime.

As \(\abs{\phi}\) increases, long-range electrostatic interactions with
protons within the Debye radius (Coulomb drag) become increasingly
important at subsonic relative velocities, as shown by the solid lines
in Figure~\ref{fig:drag-components}).
However, the Coulomb drag has a peak when \(w\) is equal to the
thermal speed of the colliders, which is
\(\approx \SI{10}{km.s^{-1}}\) for protons, giving a maximum strength from
equation~\eqref{eq:ds79} of
\begin{equation}
  \label{eq:fdrag-maximum}
  f_{\mathrm{max}} = 0.5\, (\ln\Lambda)\, \phi^2 f_* \approx 10\, \phi^2 f_* \ , 
\end{equation}
where \(\Lambda\) is the plasma parameter (number of particles within a
Debye volume), such that
\(\ln\Lambda = 23.267 + 1.5 \ln T_4 - 0.5 \ln n\).  At highly super-thermal
speeds, the Coulomb drag falls asymptotically as
\(f\drag \propto 1 / w^{2}\).  The thermal speed of electrons is higher than
that of the protons by a factor of \((m_p / m_e)^{1/2}\), so that the
electron Coulomb drag (solid light blue line) gives a second peak of
similar strength, but at \(w \approx \SI{430}{km.s^{-1}}\).  The behavior of
\(f\drag\) in all these different regimes is summarised in
Table~\ref{tab:fdrag-regimes}, in terms of \(\phi\) and
\(w_{10} = w / \SI{10}{km.s^{-1}}\).  This is further illustrated in
Figure~\ref{fig:drag-v-phi-plane}, where each of the drag regimes is
located on the \((w, \abs{\phi})\) plane.




\begin{figure}
  \centering
  \includegraphics[width=\linewidth]{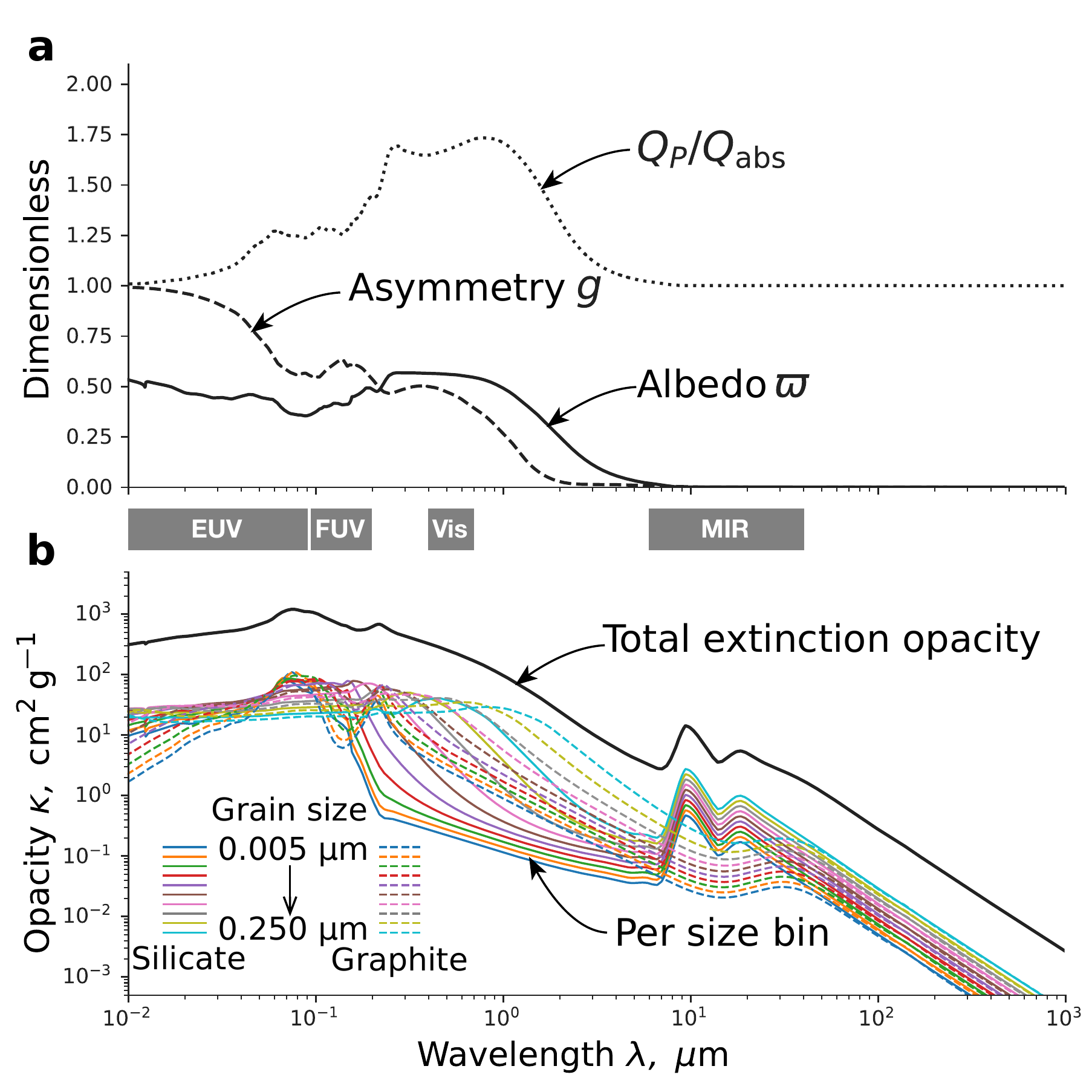}
  \caption{Extinction properties of Cloudy's standard ``ISM'' dust
    mixture. %
    (a)~Wavelength dependence of mean values over the entire mixture
    of three dimensionless quantities related to scattering: albedo,
    \(\varpi\) (solid line); scattering asymmetry,
    \(g = \langle \cos\theta \rangle\) (dashed line); ratio of radiation pressure
    efficiency to absorption efficiency, \(Q_P / Q_{\text{abs}}\)
    (dotted line).
    (b)~Wavelength dependence of mass opacity (cross section per unit
    mass of gas) for the whole mixture (heavy black line) and broken
    down by size bin and grain composition (colored lines, see key). }
  \label{fig:cloudy-ism-dust-opacity}
\end{figure}

\begin{table*}
  \centering
  \caption{Stellar parameters for example stars}
  \label{tab:stars}
  \begin{tabular}{l S S S S S S S S S}
    \toprule
    & {\(M / \si{M_\odot}\)} & {\(L_4\)}
    & {\(\dot{M}_{-7}\)} & {\(V_3\)} & {\( \eta\wind \)}
    & {Sp.~Type} 
    & {\(T_{\text{eff}} / \si{kK}\)} & {\(\lambda_{\text{eff}}\) / \si{\um}}
    & {\(S_{49}\)} 
    \\
    \midrule
    & 10 & 0.63 & 0.0034 & 2.47 & 0.0066 & {B1.5\,V} & 25.2 & 0.115 & 0.00013
                   \\
    Main-sequence OB stars
    & 20 & 5.45 & 0.492 & 2.66 & 0.1199 & {O9\,V} & 33.9 & 0.086 & 0.16
                    \\
    & 40 & 22.2 & 5.1 & 3.31 & 0.4468 & {O5\,V} & 42.5 & 0.068 & 1.41
                   \\[\smallskipamount]
    Blue supergiant star
    & 33 & 30.2 & 20.2 & 0.93 & 0.3079 & {B0.7\,Ia} & 23.5 & 0.123 & 0.016
                   \\
    \bottomrule
  \end{tabular}
\end{table*}

\subsection{Grain charging and gas--grain coupling}
\label{sec:cloudy-models-dust}

We calculate models of the physical properties of dust grains using
the plasma physics code Cloudy \citep{Ferland:2013a, Ferland:2017a},
which self-consistently solves the multi-frequency radiative transfer
together with thermal, ionization, and excitation balance of all
plasma constituents.  Cloudy incorporates grain charging as described
in \citet{Baldwin:1991a} and \citet{van-Hoof:2004a} with photoelectric
emission theory from \citet{Weingartner:2001b} and
\citet{Weingartner:2006a}.  We use the default ``ISM'' dust mixture
included in Cloudy, which comprises ten size bins each for spherical
silicate and graphite grains in the range \num{0.005} to
\SI{0.25}{\um}, and which is designed to reproduce the average
Galactic extinction curve \citep{Weingartner:2001a, Abel:2008a}.  The
optical properties of each grain species are calculated using Mie
theory \citep{Bohren:1983a}, assuming solid spheres.  The resultant
wavelength-dependent extinction properties of the mixture are
summarised in Figure~\ref{fig:cloudy-ism-dust-opacity}.

To ascertain the expected variation in grain properties in the
circumstellar environs of luminous stars, we calculate a series of
spherically symmetric, steady-state, constant density Cloudy
simulations, illuminated by the stars listed in Table~\ref{tab:stars},
which are the same ones as used in Paper~I.\@ Stellar spectra are
taken from the OSTAR2002 and BSTAR2006 grids, calculated with the
TLUSTY model atmosphere code \citep{Lanz:2003a, Lanz:2007a}.
Simulations are run for hydrogen densities of \numlist{1;10;100;e3;e4}
\si{cm^{-3}} and assuming standard \hii{} region gas phase abundances.
The calculation is stopped when the ionization front is reached and
the inner radius is chosen to be roughly 1\% of this.

\begin{figure*}
  \includegraphics[width=\linewidth]{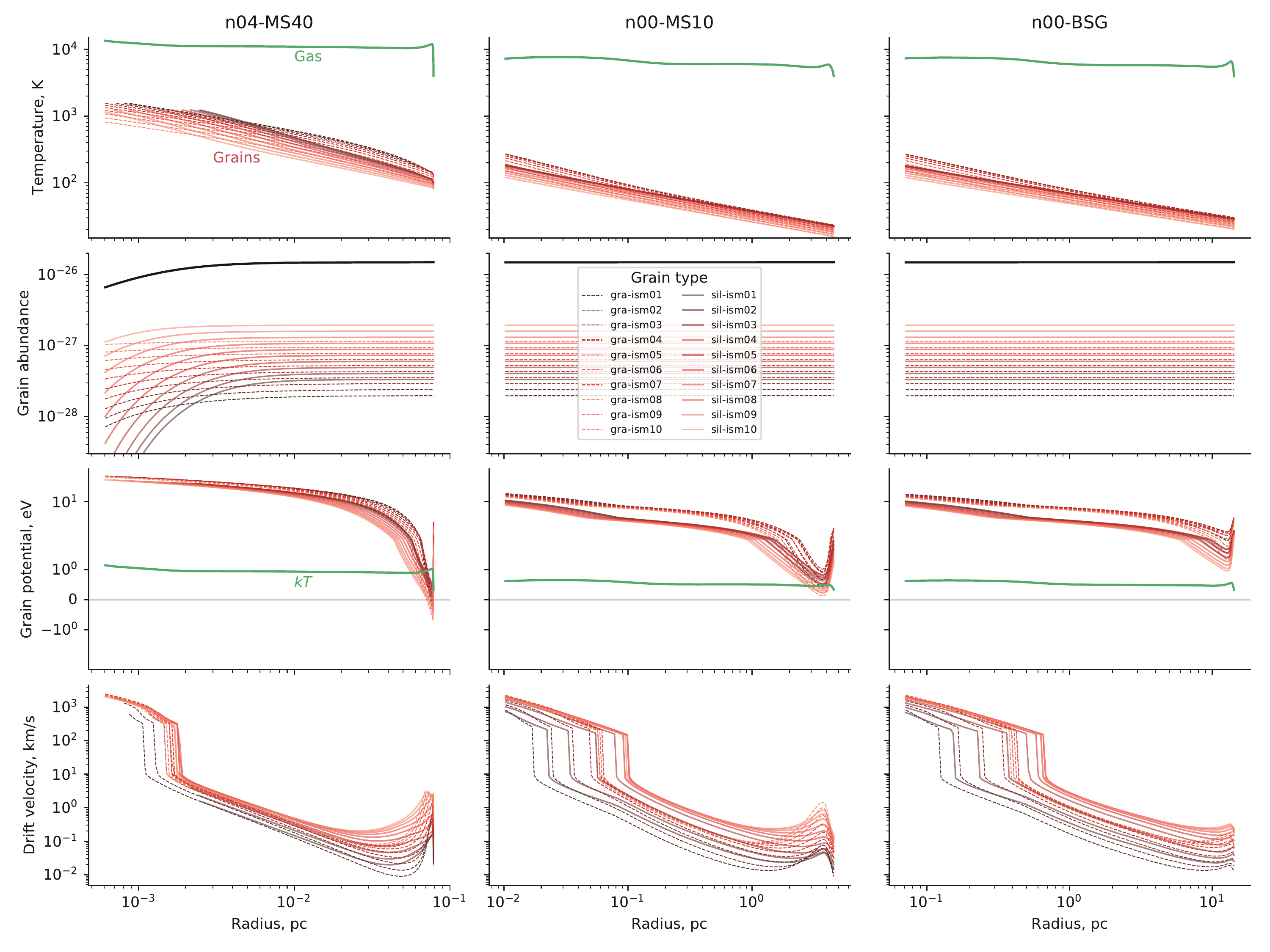}
  \caption{Dust properties as a function of radius from star for three
    selected Cloudy simulations. (a)~\SI{40}{M_\odot} main-sequence star in
    medium of density \SI{e4}{cm^{-3}}. (b)~\SI{10}{M_\odot} main-sequence
    star in medium of density \SI{1}{cm^{-3}}. (c)~Blue supergiant
    star in medium of density \SI{1}{cm^{-3}}}.
  \label{fig:multi-dustprops}
\end{figure*}

Figure~\ref{fig:multi-dustprops} shows resultant radial profiles of
dust properties for representative simulations: grain temperature, grain
abundance, grain potential, and grain drift velocity.  Line types
correspond to the different size bins of graphite and silicate grains,
as indicated in the key from smallest to largest. The left hand panels
show results for a high-density (\(n = \SI{e4}{cm^{-3}}\)), compact
(\(R \approx \SI{0.1}{pc}\)) region around an early O~star, where the grain
temperature is very high, especially for the smaller silicate grains,
and sublimation significantly reduces the grain abundance in the inner
regions \citep{Arthur:2004a}.  The remaining columns show low-density
(\(n = \SI{1}{cm^{-3}}\)), extended (\(R \sim \SI{10}{pc}\)) regions
around main-sequence and supergiant B-type stars, in which the grain
temperatures are much lower, ranging from \SIrange{20}{50}{K} in the
outer parts up to \SIrange{100}{200}{K} in the inner parts.

Unlike the strong differences in thermal properties, the radial
dependence of grain electrostatic potential (third row in
Fig.~\ref{fig:multi-dustprops}) is qualitatively similar for all the
simulations.  The grains are predominantly positively charged, with high
potentials (\(> 10\) times the thermal energy of gas particles) close
to the star due to the strong EUV and FUV photo-ejection.  The
potential falls to much lower values in the outer ionized region, as
the EUV flux falls off, and then climbs again at the ionization front
due to the fall in electron density, while the FUV photo-ejection
persists well into the neutral region.  There are small differences
between the simulations due to the increasing relative importance of the
EUV radiation for hotter stars, which leads to a deeper dip in the
potential just inside the ionization front for the \SI{40}{M_\odot} case,
even reaching negative values for some grain species.

Equilibrium drift velocity for each grain species is calculated in the
Cloudy simulations using the same \citet{Draine:1979a} theory as
described in \S~\ref{sec:drag-force-grains} and
Appendix~\ref{sec:equat-moti-grains}.  The way that this is
implemented by default in Cloudy means that if the only solution at
the inner radius is a superthermal one, then the superthermal solution
branch is followed as far as possible through the outer spatial zones.
We have modified the code so as to instead always prefer the slower
subthermal branch whenever multiple solutions are available.  This
makes more sense than the default behavior for our context, where the
grains are moving towards the star and so the radiative force is
gradually increasing from an initial low value.

Example results are shown in the bottom row of
Figure~\ref{fig:multi-dustprops} and again they are qualitatively
similar for all the simulations.  Close to the star, the radiation
force is higher than the upper limit on the Coulomb drag force
(eq.~[\ref{eq:fdrag-maximum}]), so that the equilibrium drift velocity
is exceedingly high.  We discuss this situation in greater detail
below in \S~\ref{sec:drag-force-grains}.  Note that such high drift
velocities are much higher than any realistic true relative velocity
between grains and gas, since they are based on the assumption that
the radiation force remains constant while the grain is accelerated,
which is not the case under these conditions.  Instead, they are
simply an indication that the gas and grains have completely
decoupled.

As the radial distance from the star increases, the radiation field is
increasingly diluted but the grain potential falls only slowly, so
eventually a point is reached where an equilibrium between Coulomb
drag and radiation force can be established, which corresponds to a
discontinuity in the drift velocity.  The drift velocity carries
on falling towards the outside of the \hii{} region, but then
increases again just inside the ionization front due to the drop in
grain potential there.

\subsection{Gas--grain separation: drift and rip}
\label{sec:gas-grain-separ}

In \S~\ref{sec:gas-free-bow} we calculate the behaviour of an
incoming stream of dust grains, subject only to the repulsive
radiation force from a star.  For an initial inward radial trajectory,
the dust grain motion is decelerated and turned around, reaching a
minimum radius \(R\starstar\), given by equation~\eqref{eq:dust-r0}.
This drag-free radiative turnaround radius, \(R\starstar\), is smaller
for higher initial inward velocities, but is independent of the
density of the incoming stream.  We are now in a position to see how
gas--grain drag will modify this picture.

We introduce the local radiation parameter, \(\Xi\), defined as the
ratio of direct stellar radiation pressure to gas pressure:
\begin{equation}
  \label{eq:Xi-Prad-over-Pgas}
  \Xi \equiv \frac{P\rad}{P\gas} \approx \frac{L}{4 \pi R^2 c\, (2 n k T)} \ ,
\end{equation}
where the last expression corresponds to the optically thin limit.  If
we define the grain's frequency-averaged radiation pressure efficiency
as
\begin{equation}
  \label{eq:Qpbar}
  \Qpbar = \frac{1}{L} \int_0^\infty \!\Qp \, L_\nu \, d\nu \ ,
\end{equation}
then equations~\eqref{eq:dust-rad-force} and~\eqref{eq:fstar} give the
radiation force acting on a grain as
\begin{equation}
  \label{eq:frad-Xi}
  f\rad = \Qpbar\, \Xi\, f_* \ .
\end{equation}

\begin{figure}
  \centering
  \includegraphics[width=\linewidth]{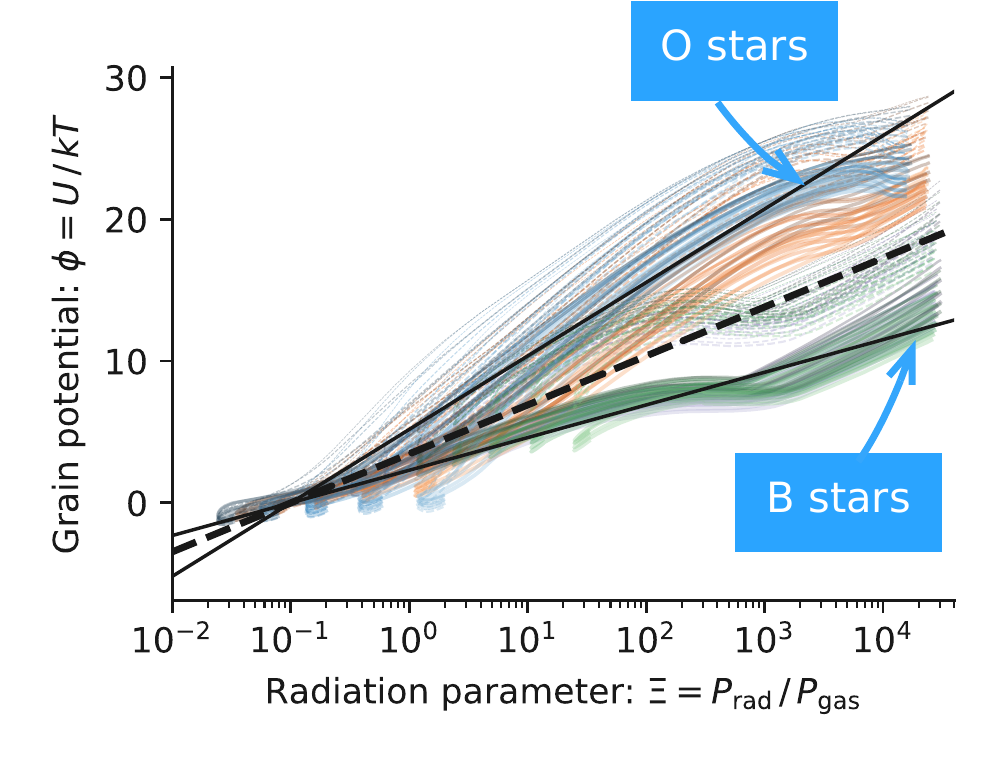}
  \caption{Grain potential in thermal units (linear scale) versus
    radiation parameter (logarithmic scale). All densities and stellar
    types are shown, with line colors as indicated (blue/orange for
    O~stars, purple/green for B~stars) and lighter shades indicating
    higher gas densities.  Solid lines show silicate grains and dashed
    lines show graphite grains.  Line width increases with grain size
    (to reduce clutter, only every second size bin is shown).
    Straight black lines show the logarithmic fits discussed in the
    text: eq.~\eqref{eq:phi-vs-Xi}, most appropriate for carbon grains
    around cooler stars, is shown by the dashed line, while the solid
    lines show the modifications for silicate grains and for hotter
    stars.}
  \label{fig:phi-vs-Xi}
\end{figure}

\begin{figure*}
  \includegraphics[width=\linewidth]{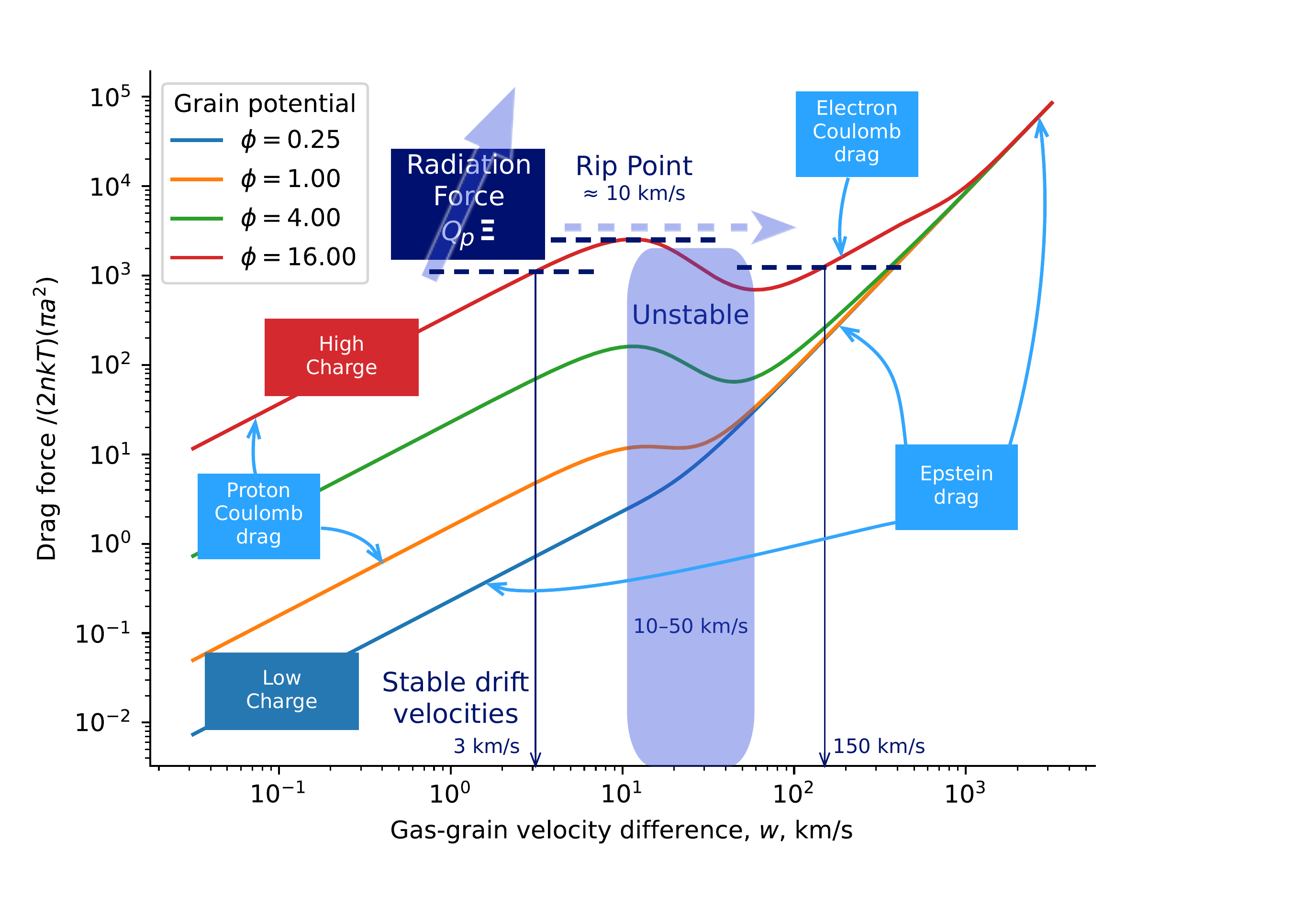}
  \caption{Dimensionless drag force, \(f\drag / f_*\), as a function
    of gas--grain slip velocity, \(w\), for different values of the
    grain potential in thermal units, \(\phi\).  Contributions from
    proton and electron Coulomb (electrostatic) drag, as well as
    Epstein (solid-body) drag are indicated.  Examples of subsonic and
    highly supersonic stable drift velocities are shown (thin dark
    blue arrows), where the drag force is in equilibrium with the
    radiation force (thick dark blue dashed lines), while blue shading
    indicates the unstable, mildly supersonic velocity regime, where
    no stable drift equilibrium exists. }
  \label{fig:gas-grain-drag-photoionized}
\end{figure*}

The grain potential \(\phi\), which is crucial for determining
\(f\drag\) (Tab.~\ref{tab:fdrag-regimes}), is due to competition
between the photons and the charged particles that interact with the
grain.  It is therefore reasonable to suppose that \(\phi\) should also
be primarily determined by \(\Xi\).  This is confirmed in
Figure~\ref{fig:phi-vs-Xi} using the Cloudy simulations of
\S~\ref{sec:cloudy-models-dust}, for which we find a slow dependence
that can be approximated as
\begin{equation}
  \label{eq:phi-vs-Xi}
  \phi(\Xi) \approx 1.5 \bigl( 2.3 +  \ln \Xi \bigr) \ .
\end{equation}
There are also slight secondary dependencies on the grain composition and
stellar spectrum.  The relationship given in eq.~\eqref{eq:phi-vs-Xi}
is appropriate for graphite grains and for stellar effective
temperatures in the range \SIrange{20}{30}{kK}.  For hotter stars than
this, \(\phi\) should be multiplied by a further factor of \(1.5\), while
for silicate grains it should be divided by \(1.5\).

In the outer regions of the photoionized volume around an OB star,
close to the ionization front, the radiation parameter is low, with
typical value \(\Xi \sim 0.1\).  In this regime, the negative charge
current at the grain surface due to electron collisions is roughly in
balance with the positive current due to the ultraviolet photoelectric
effect \citep{Weingartner:2001b}, leading to a low grain potential,
\(\abs{\phi} < 1\), which may be positive or negative.  The low
\(\Xi\) means that the radiative force is also weak:
\(f\rad \sim 0.1 f_*\) from equation~\eqref{eq:frad-Xi} if
\(\Qpbar \approx 1\) at UV wavelengths, which is true for all but the smallest
grains.  Thus, from the equations for \(f\drag\) given in
Table~\ref{tab:fdrag-regimes}, the radiative force can be balanced by
Epstein drag if \(w_{10} \sim 0.1\), leading to a small equilibrium drift
velocity, \(w\drift < \SI{1}{km.s^{-1}}\), of the grains with respect
to the gas.  This drift is much smaller than the inward stream
velocities that we are considering
(\(v_\infty > \SI{10}{km.s^{-1}}\)), so the dust follows the gas stream at
a slightly reduced velocity (\(< 10\%\)), and (by mass conservation) a
slightly increased density.  Each grain exerts an exactly opposite
force to \(f\drag\) upon the gas, but since the dust-gas mass ratio,
\(Z\grain\), is small, this produces a negligible acceleration of the
gas.

\begin{table}
  \caption{Critical values of radiation parameter at the rip point: \(\Xi_\dag\)}
  \centering
  \begin{tabular*}{0.75\columnwidth}{l @{\quad\quad\quad\quad} S S} \toprule
    & \multicolumn{2}{c}{Grain composition} \\
    Spectrum & {Graphite} & {Silicate}
    \\ \midrule
    B star & 1000 +- 400 & 350 +- 150 \\
    O star & 3000 +- 500 & 2500 +- 500 \\
    \bottomrule
    \addlinespace
    \multicolumn{3}{@{}p{0.75\columnwidth}@{}}{
    Calculated from the Cloudy models shown in Figure~\ref{fig:drift-gn}. 
    Uncertainties represent variations with grain size and gas density. 
    }
  \end{tabular*}
  \label{tab:Xi-rip}
\end{table}

As the dusty stream approaches the star, the radiation parameter
\(\Xi\) will increase, with a dependence of \(R^{-2}\) once the stream
is well inside the ionization front, assuming roughly constant
pressure in the \hii{} region.  This increases \(f\rad\)
(eq.~[\ref{eq:frad-Xi}]), but also increases the grain potential
\(\phi\) (eq.~[\ref{eq:phi-vs-Xi}]) due to the increasing dominance of
grain charging by photoelectric ejection.  Initially, this results in
a lowering of the equilibrium drift velocity to \(w_{10} \sim 0.01\) as
the Coulomb drag kicks in (see lower panels of
Fig.~\ref{fig:multi-dustprops}).  However, at smaller radii the slow
logarithmic increase in \(\phi(\Xi)\) means that the drift velocity must
start increasing again to accommodate the linear increase of
\(f\rad(\Xi)\).  Eventually, \(f\rad\) exceeds \(f_{\mathrm{max}}\), the
maximum drag force that proton Coulomb interactions can provide
(eq.~[\ref{eq:fdrag-maximum}]).  This occurs at a critical value of
the radiation parameter, which we denote the \textit{rip point}:
\(\Xi_\dag \sim 1000\). The process is illustrated in
Figure~\ref{fig:gas-grain-drag-photoionized}, which shows the regions
of stability and instability as a function of gas--grain slip velocity
as the grain potential and radiation force are increased.

\begin{figure*}
  \includegraphics[width=\linewidth]{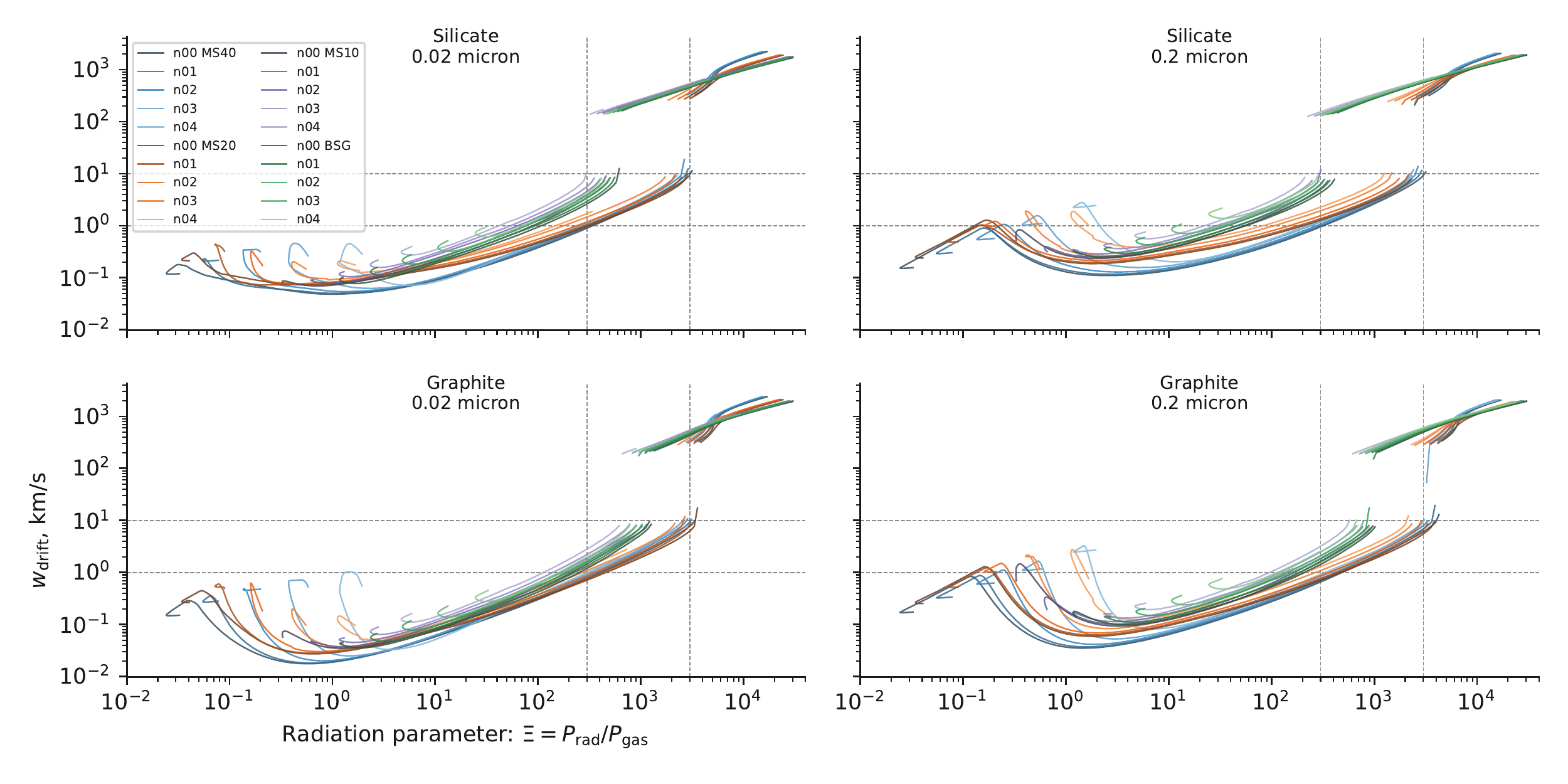}
  \caption{Drift velocity \(w\drift\) versus radiation parameter
    \(\Xi\). Each line represents a simulation with ambient density and
    stellar type as indicated in the key.  Results are shown for
    graphite and silicate grains of two different sizes.  The rip
    point, which corresponds to gas--grain decoupling, is the
    discontinuity in the curves at
    \(w\drift \approx \SI{10}{km.s^{-1}}\), indicated by the upper
    horizontal dashed line.  The vertical dashed lines show the narrow
    range of radiation parameter, \(\Xi = 1000 \pm \SI{0.5}{dex}\), that
    encompasses the rip point for all simulations. }
  \label{fig:drift-gn}
\end{figure*}

To test these ideas, we plot in Figure~\ref{fig:drift-gn} the grain
drift velocity from the Cloudy simulations as a function of \(\Xi\),
showing results for four different grain types and for all
combinations of stellar parameters and ambient densities for which we
have run simulations.  It can be seen that the radiation parameter at
the rip point \(\Xi_\dagger\) is indeed confined to a narrow range.  The
fundamental explanation for this is that both the charge balance and
the force balance are essentially due to competition between the
photons and the charged particles that interact with the grain.  The
small variations in \(\Xi_\dag\) with stellar type and grain composition,
which are of order \SI{+- 0.5}{dex}, are listed in
Table~\ref{tab:Xi-rip}.  The gas density and grain size have very
little influence on this critical value \(\Xi_\dag\), with the only
exception being the very smallest grains (\(a < \SI{0.006}{\um}\), not
illustrated), which show \(\Xi_\dag \approx \num{e4}\), but such grains are only
minor contributors to the UV opacity for our adopted dust mixture
(\(< 10\%\) in EUV and \(< 1\%\) in FUV, see
Fig.~\ref{fig:cloudy-ism-dust-opacity}).

The radius of the rip point, \(R_\dag\), can be expressed in terms of
\(R_*\), the fiducial optically thick bow shock radius
(eq.~[\ref{eq:Rstar}]):
\begin{equation}
  \label{eq:Rdag-over-Rstar}
  R_\dag = \frac{v_\infty}{\sound}\, \Xi_\dag^{-1/2} R_* \approx v_{10}\, \Xi_\dag^{-1/2} R_* \ ,
\end{equation}
where we have made use of equation~\eqref{eq:Xi-Prad-over-Pgas} and
the final approximate equality assumes a typical \hii{} region
temperature of \SI{e4}{K}.

\begin{figure*}
  \centering
  \includegraphics[width=\linewidth]{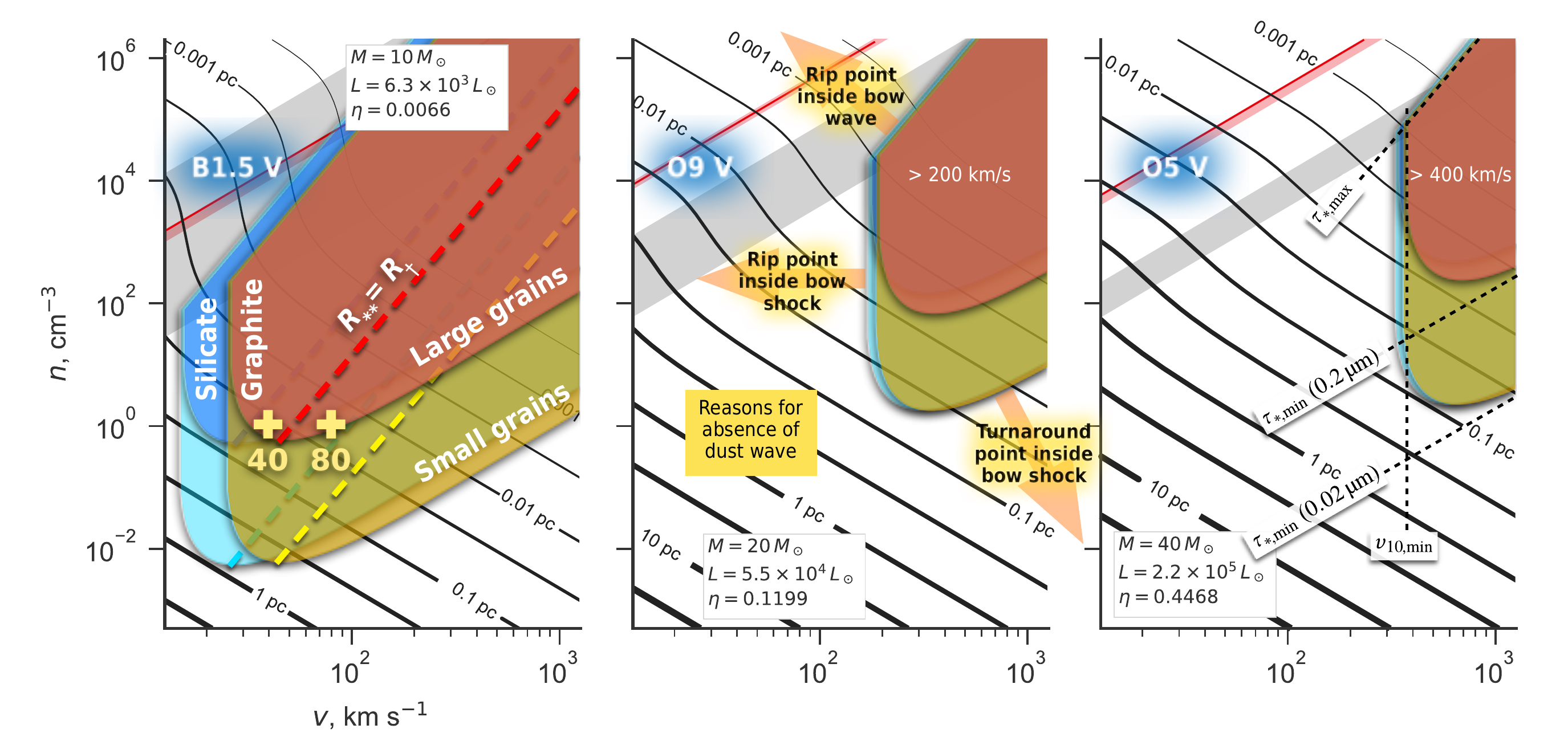}
  \caption{Regions of stream parameter space \((v, n)\) where dust
    waves may form around main-sequence OB stars of
    \SIlist{10;20;40}{M_\odot} with standard wind parameters (see
    Tab.~\ref{tab:stars}).  This figure is similar to Paper~I's
    Fig.~2, except that the velocity axis is logarithmic and extends
    out to \SI{1000}{km.s^{-1}}.  Overlapping colored shapes show
    parameters where dust waves may be allowed in the cases of large
    (\(a = \SI{0.2}{\um}\)) and small (\(a = \SI{0.02}{\um}\))
    graphite and silicate grains, as labeled in the left panel.  For
    \((v_\infty, n)\) outside of these shapes, dust waves cannot occur for
    the reasons indicated by labeled orange arrows in the center
    panel.  Labeled dashed lines in the right panel show the
    correspondence between the region boundaries and each dust wave
    existence condition given in
    equations~(\ref{eq:dust-wave-velocity-condition},
    \ref{eq:dust-wave-low-density-condition},
    \ref{eq:dust-wave-high-density-condition}). Heavy dashed lines in
    the left panel show where the rip point and the drag-free
    turnaround radius coincide.  Dust waves above and to the left of
    these lines are drag confined, while dust waves below and to the
    right of the lines are inertia confined.  Yellow plus symbols show
    the parameters for the two example trajectories shown in
    Fig.~\ref{fig:phase-space-trajectories}.}
  \label{fig:existence-dust-wave}
\end{figure*}

\subsection{Existence conditions for dust waves}
\label{sec:exist-cond-separ}

In order for a separate outer dust wave to exist, it is necessary for
the grains to decouple from the incoming gas stream before the stream
hits the hydrodynamic bow shock caused by the stellar wind.  The wind
bow shock radius is \(R_0 = \eta\wind^{1/2} R_*\) (eq.~[\ref{eq:x-cases}]),
where \(\eta\wind\) is the wind momentum efficiency
(eq.~[\ref{eq:wind-eta-typical}]).
Therefore, the condition
\(R_\dag > R_0\) becomes from equation~\eqref{eq:Rdag-over-Rstar}:
\begin{equation}
  \label{eq:dust-wave-velocity-condition}
  v_{10} > v_{10,\text{min}} = \bigl( \Xi_\dag \, \eta\wind \bigr)^{1/2} \ . 
\end{equation}
For early O main-sequence stars and OB supergiants, the wind
efficiency is generally high (\(\eta\wind > 0.1\)) and
\(\Xi_\dag > 2000\) (Tab.~\ref{tab:Xi-rip}), so that dust waves can only
exist when the stream velocity is very high
(\(v_\infty > \SI{150}{km.s^{-1}}\)).  For main-sequence B~stars, in
contrast, the wind can be much weaker (\(\eta\wind < 0.01\)) and
\(\Xi_\dag\) is also smaller, so that dust waves are permitted by this
criterion for much lower stream velocities:
(\(v_\infty > \SI{30}{km.s^{-1}}\)).  The same will be true of the
``weak-wind'' class of late O main-sequence stars, which also show
\(\eta\wind < 0.01\) (see Paper~I, \S~4).

However, there are other conditions that need to be satisfied in order
for the dust wave to exist.  For instance, the drag-free turnaround
radius must also be outside the bow shock: \(R\starstar > R_0\),
otherwise the radiation is incapable of repelling the grain
opportunely, even once it has decoupled from the gas.  From
equation~\eqref{eq:dust-r0}, together with Paper~I's
equations~(\ref{eq:Rstar}, \ref{eq:tau-star}),
we find
\begin{equation}
  \label{eq:Rstarstar-over-Rstar}
  \frac{R\starstar}{R_*} = \frac{2 \kappa\grain \tau_*}{\kappa} \ , 
\end{equation}
so the condition becomes
\begin{equation}
  \label{eq:dust-wave-low-density-condition}
  \tau_* >  \tau_{*,\text{min}} = 0.5\, \frac{\kappa}{\kappa\grain}\, \eta\wind^{1/2} 
  \ . 
\end{equation}
The average value of the factor \(\kappa / \kappa\grain\) over the entire grain
population must be equal to the dust--gas mass ratio,
\(Z\grain \approx 0.01\), but the factor will vary between grains, according
to their size and composition.\footnote{%
  Recall that \(\kappa\) is the opacity per unit mass of gas, while
  \(\kappa\grain\) is the opacity per unit mass of a particular grain. In
  both cases, averaged over the stellar spectrum.} %
In particular, it will be relatively larger for the largest grains
(\(a \approx \SI{0.2}{\um}\)), which dominate the total dust mass, and
smaller for the smaller grains (\(a \approx \SI{0.02}{\um}\)), which
dominate the UV opacity.  Given the dependence of \(\tau_*\) on the
stream parameters (eq.~[\ref{eq:taustar-typical}]),
for a given
stellar luminosity this condition corresponds to a minimum value for
\(n / v_\infty^2\).

A third condition comes from requiring \(R_\dag > R_0\) in the radiation
bow wave regime (see Paper~I's \S~2.1),
where
\(R_0 \approx 2 \tau_* R_*\).  This yields
\begin{equation}
  \label{eq:dust-wave-high-density-condition}
  \tau_* < \tau_{*,\text{max}} = 0.5\, v_{10}\, \Xi_\dag^{-1/2} \ , 
\end{equation}
which, for a given stellar luminosity, corresponds to a maximum value
for \(n / v_\infty^4\).  Thus, for a given stream velocity that satisfies
equation~\eqref{eq:dust-wave-velocity-condition},
equations~(\ref{eq:dust-wave-low-density-condition},
\ref{eq:dust-wave-high-density-condition}) determine respectively the
minimum and maximum stream density for which a dust wave can exist.   

The combined effects of the three conditions are illustrated in
Figure~\ref{fig:existence-dust-wave} for each of the three example
main sequence stars from Table~\ref{tab:stars}.  Further restrictions
on the existence of dust waves arise when the effects of magnetic
fields are considered, as will be discussed in
\S~\ref{sec:magn-effects-grain} below.  Note that the three conditions
are restrictions solely on the formation of an \textit{outer} dust
wave, that is, outside of the wind-supported hydrodynamic bow shock.
In the case of the equation~\eqref{eq:dust-wave-low-density-condition}
condition, there is a further possibility: if the gas--grain coupling
(and magnetic coupling) is so weak that it is still unimportant at the
higher densities found in the bow shock shell, then an
inertia-confined \textit{inner} dust wave may form inside the bow
shock, even when \(\tau_* < \tau_{*,\text{min}}\).  This is similar to the
case studied by \citet{Katushkina:2017a, Katushkina:2018a}, which
requires simultaneous modeling of the grain dynamics with the
magnetohydrodynamics of the bow shock.  Note that unlike with
equation~\eqref{eq:dust-wave-low-density-condition}, violation of
equation~\eqref{eq:dust-wave-velocity-condition} cannot lead to an
inner dust wave, since the density compression in the bow shock will
reduce the radiation parameter, \(\Xi\), which moves the rip point,
\(R_\dag\), to an even smaller radius.  Therefore, if radiation has not
managed to decouple a grain before it passes through the shock, it is
unlikely to be able to do so afterwards.

\begin{figure*}
  \centering
  \includegraphics[width=\linewidth]{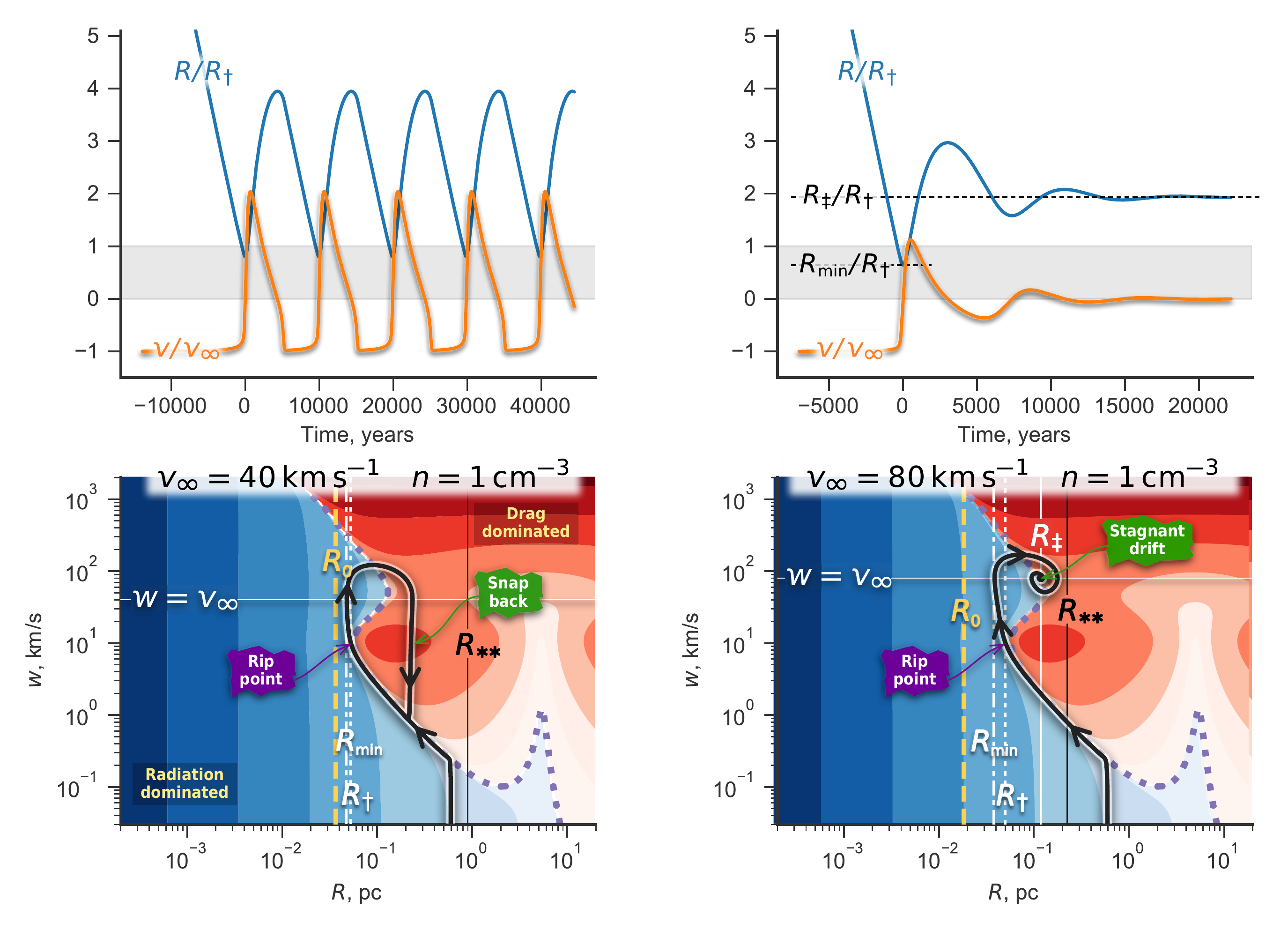}
  \caption{Trajectories of small graphite grains
    (\(a = \SI{0.02}{\um}\)) at impact parameter \(b = 0\) for two
    example cases (see yellow ``+'' symbols in left panel of
    Fig.~\ref{fig:existence-dust-wave}), which differ only in the
    stream velocity: \(v = \SI{40}{km.s^{-1}}\) (left panels) and
    \SI{80}{km.s^{-1}} (right panels).  In both cases, the stream
    density is \(n = \SI{1}{cm^{-3}}\) and the central star is a
    \SI{10}{M_\odot} main-sequence B star (see Tab.~\ref{tab:stars}).
    Upper panels show the evolution of the grain position, \(R\) (blue
    curve, normalized by the rip point radius, \(R_\dag\)), and grain
    velocity, \(v\) (orange curve, normalized by the gas stream
    velocity).  The origin of the time axis is set to the moment of
    closest approach of the grain to the star: \(R = \Rmin\).  Lower
    panels show the trajectories in phase space: position versus
    gas--grain relative slip velocity (\(w = \abs{v - v_\infty}\)).  Filled
    contours show the net force on the grain: \(f\rad - f\drag\), with
    positive values in blue and negative values in red.  The heavy
    dotted line shows where there is no net force: \(f\rad = f\drag\).
    The grain trajectory (thick, solid black line with arrows)
    initially follows this line, but departs from it after the rip
    point. In the left panel, the grain enters a limit cycle between
    decoupling (rip) and re-coupling (snap back).  In the right panel,
    the grain spirals in on the stagnant drift point.  See text for
    further details.}
    \label{fig:phase-space-trajectories}
\end{figure*}

\subsection{Post-rip grain dynamics}
\label{sec:two-regimes-post}

We now investigate the trajectory of the dust grain following the
catastrophic breakdown of gas--grain coupling at the rip point.  Two
regimes are possible, depending on the relation between the rip point
radius, \(R_\dag\), and the drag-free radiative turnaround radius,
\(R\starstar\).  If \(R_\dag > R\starstar\), then the grain's inertia
will still carry it in as far as \(R\starstar\) and the initial
trajectory will be almost identical to that described in
\S~\ref{sec:gas-free-bow} for the drag-free case. But once the grain
has been turned around by the radiation field and pushed out past
\(R_\dag\) again, it will \emph{recouple} to the gas.  
We will refer to this as an \textit{inertia-confined dust wave} (IDW).
From equations~(\ref{eq:Rdag-over-Rstar},
\ref{eq:Rstarstar-over-Rstar},
\ref{eq:dust-wave-high-density-condition}), the condition
\(R_\dag > R\starstar\) corresponds to
\begin{equation}
  \label{eq:IDW}
  \tau_* < \frac{\kappa}{\kappa\grain} \tau_{*,\text{max}} \ , 
\end{equation}
which is indicated by dashed lines in the left panel of
Figure~\ref{fig:existence-dust-wave}.  If, on the other hand,
\(R_\dag < R\starstar\), then the tail wind provided by the gas carries
the grain closer to the star than its inertia would naturally take it.
When the grain finally decouples at \(R_\dag\) it experiences a much
higher unbalanced \(f\rad\), which can initially accelerate it to
outward velocities significantly higher than the inflow velocity if
\(R_\dag \ll R\starstar\).  We will refer to this case as a
\textit{drag-confined dust wave} (DDW).  As in the IDW case, the
expelled grain will eventually recouple to the gas once it moves away
from the star.

\label{sec:grain-traj-along}

What happens to the grain after recoupling depends on the sign of
\(d f\drag / d w\) when \(w = \abs{v_\infty}\).  If this derivative is
positive, as is the case in drag regimes~II and V (see
Tab.~\ref{tab:fdrag-regimes} and Fig.~\ref{fig:drag-v-phi-plane}),
then the grain can reach a stable equilibrium drift at rest with
respect to the star at a point \(R_\ddag\), which we call the
\textit{stagnant drift radius}. If the stream velocity is not
excessively high (\(v_\infty < \SI{150}{km.s^{-1}}\) when
\(\phi = 4\), or \(< \SI{300}{km.s^{-1}}\) when \(\phi = 16\)), then the
equilibrium \(f\rad\) is less than the value at the rip point,
requiring a lower value of the radiation parameter:
\(\Xi_\ddag < \Xi_\dag\).  The resultant stagnant drift radius is therefore
outside the rip point: \(R_\ddag > R_\dag\).  Of course, a static equilibrium
is only possible when the impact parameter is exactly zero.
Otherwise, there will be an unbalanced lateral component of the
radiation force, which will cause a sideways drift.  However, as we
show below, strong coupling to the magnetic field means that the
strictly on-axis calculation is a reasonable approximation over a
range of impact parameters in the case where the angle between the
magnetic field direction and the stream velocity is not too large.

On the other hand, if \(d f\drag / d w < 0\) when
\(w = \abs{v_\infty}\), then the equilibrium is unstable and no stagnant
drift is possible.  This occurs for drag regime~IV, which applies when
\(\phi > 1\) and
\(\SI{10}{km.s^{-1}} < v_\infty < \SI{50}{km.s^{-1}}\), as illustrated in
Figure~\ref{fig:gas-grain-drag-photoionized}.  There is also a second
unstable regime (partially visible in the upper-right corner of
Fig.~\ref{fig:drag-v-phi-plane}), which is related to the thermal peak
in the electron Coulomb drag when \(\phi > 30\) and
\(\SI{400}{km.s^{-1}} < v_\infty < \SI{2000}{km.s^{-1}}\).  This is not
relevant to bow shocks around OB~stars since \(\phi\) does not reach such
high values, but it may apply in other contexts, such as outflows from
AGN, where grain potentials as high as \(\phi \sim 100\) can be achieved
\citep{Weingartner:2006a}.

An example of each of these two behaviors is illustrated in
Figure~\ref{fig:phase-space-trajectories}.  The left panels show the
case where \(v_\infty = \SI{40}{km.s^{-1}}\), which is in the unstable
regime, resulting in periodic ``limit-cycle'' behavior (the parameters
of this model correspond to the yellow ``plus'' symbol labeled ``40''
in the left panel of Fig.~\ref{fig:existence-dust-wave}).  During the
grain's first approach, it starts to follow a phase trajectory (lower
left panel) along the \(f\rad - f\drag = 0\) contour, corresponding to
equilibrium drift, in which the grain begins to move a few
\si{km.s^{-1}} slower than the gas stream.  Then, when it reaches the
rip point (\(R = R_\dag\), \(w \approx \SI{10}{km.s^{-1}}\)) it suddenly
experiences a large unbalanced outward radiation force (blue region of
phase space in Fig.~\ref{fig:phase-space-trajectories}). The grain's
inward momentum carries it to the point \(\Rmin \approx 0.85 R_\dag\), before
it is expelled at roughly twice the inflow speed.  However, after
moving outward, it finds itself in a drag-dominated region of phase
space (red in the figure), and so recouples to the inflowing gas
stream.  The recoupling initiates gradually, as the grain's outward
motion is slowed and turns around, but is completed suddenly once
\(w\) again falls below \SI{10}{km.s^{-1}}, in what we term
\textit{snap back}. The net result is that the grain has returned to
exactly the same phase track that it started in on, and so repeats the
cycle indefinitely.

The right panels of Figure~\ref{fig:phase-space-trajectories} show the
case where the stream velocity is doubled to
\(v_\infty = \SI{80}{km.s^{-1}}\), but all other parameters remain the
same.  At this velocity, the equilibrium drift is stable and so the
grain can achieve a stagnant drift solution, where it is stationary
with respect to the star.  The trajectory during the first approach is
similar to the previous case, except that the overshoot of the rip
point is greater, so that \(\Rmin \approx 0.65 R_\dag\) in this case.  This is
a consequence of the fact that the rip point is closer to the
drag-free turnaround radius (\(R_\dag / R\starstar\) is larger than in
the lower velocity case), so that the grain inertia is relatively more
important.  A second consequence of this is that the speed of the
initial expulsion is not so large, being only a little higher than the
inflow velocity.  The qualitative difference between the two cases
emerges after the first recoupling: instead of the snap back and
endless limit cycle, the grain oscillates about the stagnant drift
radius with ever decreasing amplitude, so that after a few oscillation
periods it has come to almost a complete rest.

\begin{figure}
  \includegraphics[width=\linewidth]{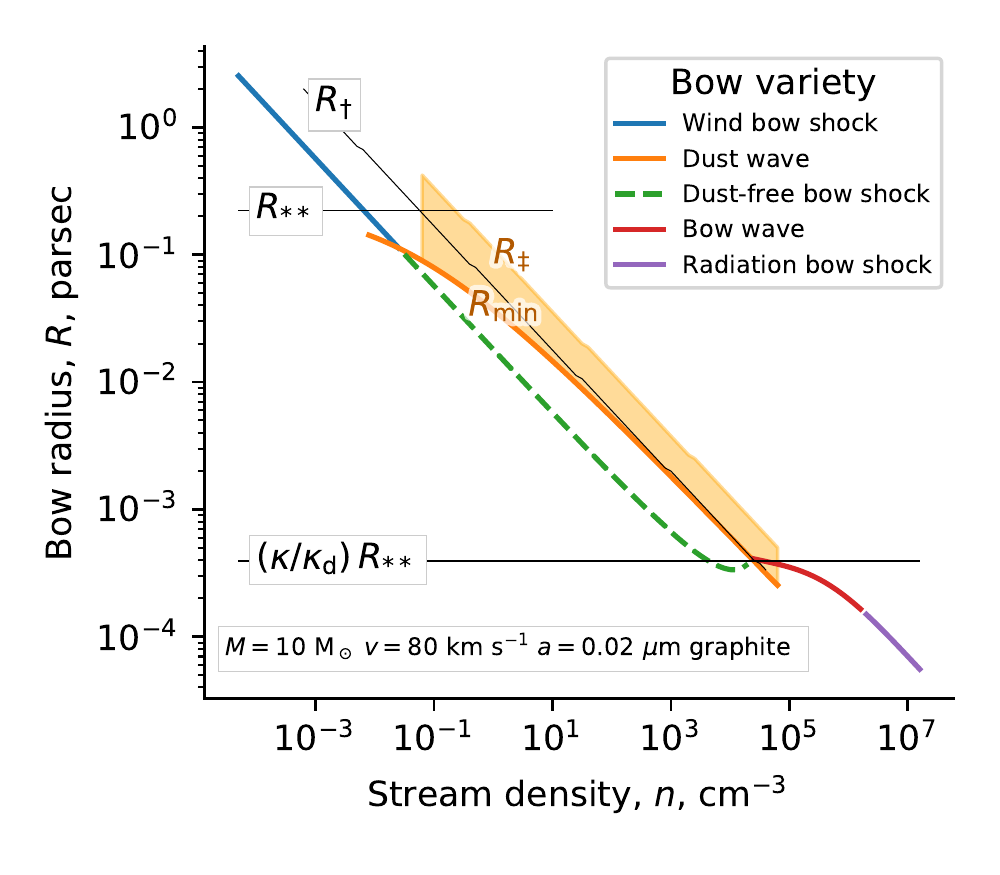}
  \caption{Bow radius as a function of stream density for a stream of
    initial velocity \SI{80}{km.s^{-1}}, which interacts with a
    \SI{10}{M_\odot} main-sequence B~star.  This corresponds to a vertical
    slice through the left panel of
    Fig.~\ref{fig:existence-dust-wave}.  At low densities, the
    hydrodynamic bow shock (blue line) is larger than the drag-free
    turnaround radius for small carbon grains, meaning that a grain's
    inertia carries it into the bow shock along with the gas, even
    though the gas--grain coupling is not particularly strong.  At
    densities above about \SI{0.05}{cm^{-3}}, however, this is no
    longer true and a separate dust wave forms outside of the
    hydrodynamic bow shock, which is now dust-free (green dashed
    line).  The grains in the dust wave will occupy a range of radii
    (pale orange shading) between \(\Rmin\) (solid orange line) and
    \(R_\ddag\), the stagnant drift radius.  At densities above about
    \SI{1000}{cm^{-3}}, the gas stream starts to feel the effect of
    passing through the dust wave, and above \SI{3e4}{cm^{-3}}, the
    dust wave and bow shock merge to form a radiative bow wave (red
    line), which becomes an optically thick radiative bow shock
    (purple line) above \SI{e6}{cm^{-3}}.}
  \label{fig:decouple-vertical-cut}
\end{figure}

\subsection{Back reaction on the gas flow}
\label{sec:back-reaction-gas}

So far we have ignored the effect of the drag force on the gas stream
itself, but it is clear that this must become important as \(\tau_*\)
approaches \(\tau_{*,\text{max}}\), since that is the point where the
dust wave transitions to a bow wave, in which the dust and gas are
perfectly coupled.  A full treatment of this problem would require
solving the hydrodynamic equations simultaneously with the equations
of motion of the dust grains, which is beyond the scope of this paper.
Instead, we outline a heuristic approach that qualitatively captures
the physics involved.

The maximum drag force experienced by a grain is at the rip point.
Since the grain follows a zero-net-force phase track up until that
point, this can be written with the help of
equations~(\ref{eq:dust-rad-force}, \ref{eq:dust-r0}) as
\begin{equation}
  \label{eq:fdrag-max}
  f\drag (R_\dag) = f\rad(R_\dag) =   \frac{m\grain v_\infty^2 R\starstar}{ 2 R_\dag^2} 
\end{equation}
The timescale of the flow can be characterized by the crossing time
\(R_\dag / v_\infty\), but the residence time of the grain at the bow apex
will be several times larger than this (see previous section).  On the
other hand, the average drag force during this residence will be
several times smaller than \(f\drag (R_\dag)\) if
\(R_\ddag > R_\dag\), which is typically the case.  We therefore parameterize
our ignorance via a dimensionless factor, \(\alpha\), which we expect to be
of order unity, and write the total impulse imparted to the grain by
drag as
\begin{equation}
  \label{eq:grain-impulse}
  J\drag \equiv \int \!f\drag \, dt \approx \alpha f\drag (R_\dag) \frac{R_\dag}{v_\infty}
  = \tfrac12 \alpha \, m\grain v_\infty \, \frac{R\starstar}{R_\dag} \ .
\end{equation}

By Newton's Third Law, an equal and opposite impulse is imparted to
the gas, which will act to decelerate the gas stream as it decouples
from the grains.  Realistically, \(J\drag\) should be summed over the
grain size distribution, but for simplicity we assume that all grains
are identical, so that the mass of gas that accompanies each grain is
given by
\begin{equation}
  \label{eq:gas-mass}
  m_{\text{gas}} = \frac{m\grain}{Z\grain} =  m\grain \, \frac{\kappa\grain}{\kappa} \ . 
\end{equation}
If the gas remains supersonic after decoupling, then thermal pressure
can be ignored and the gas will suffer a change in momentum equal to
\(J\drag\), so that its velocity is reduced by
\(\Delta v = J\drag / m_{\text{gas}}\), which by
equations~(\ref{eq:Rdag-over-Rstar}, \ref{eq:Rstarstar-over-Rstar},
\ref{eq:dust-wave-high-density-condition}, \ref{eq:grain-impulse},
\ref{eq:gas-mass}) is
\begin{equation}
  \label{eq:gas-dv}
  \Delta v = \tfrac12 \alpha \frac{\tau_*}{\tau_{*, \text{max}}} v_\infty\ .
\end{equation}
This deceleration reduces the gas stream's ram pressure before it
interacts with the central star's stellar wind.  The radius of the
dust-free bow shock formed by this interaction is therefore increased
by a factor \((1 - \Delta v / v_\infty)^{-1}\) with respect to the value given
by equation~\eqref{eq:x-cases}, yielding
\begin{equation}
  \label{eq:gas-free-bow-shock}
  R_{\text{dfbs}} \approx \frac{\eta\wind^{1/2} R_*}{1 - \tfrac12 \alpha \tau_* / \tau_{*, \text{max}}} \ .
\end{equation}

An example is illustrated in Figure~\ref{fig:decouple-vertical-cut},
where the dust-free bow shock radius is shown by the green dashed line
as a function of stream density, \(n\).  This is calculated for fixed
stream velocity and grain and star properties, so that
\(\tau_* \propto n^{1/2}\) (eq.~[\ref{eq:taustar-typical}]).
In order for
\(R_{\text{dfbs}}\) to match the dust-wave and bow-wave radii at the
point where they cross at \(\tau_* = \tau_{*, \text{max}}\), we find
\(\alpha \approx 1.5\) is required.  It can be seen that the gas deceleration is
negligible over most of the density range for which a separate dust
wave arises.  Only for \(n > \SI{e3}{cm^{-3}}\) does
\(R_{\text{dfbs}}\) begin to curve up from the general \(n^{-1/2}\)
trend, becoming essentially flat at a value
\(R_{\text{dfbs}} \approx (\kappa/\kappa\grain) R\starstar\) until full-coupling is
established at \(n > \SI{3e4}{cm^{-3}}\).  Note, however, that the
treatment described here is very approximate: it does not take into
account the shock that must form once \(J\drag\) reaches an
appreciable fraction of \(m_{\text{gas}} v_\infty\) and, additionally, it
includes a factor, \(\alpha\), whose value has not been rigorously
justified.  More detailed modeling is required to fully understand the
bow behavior in this transition regime.

\section{Magnetic coupling of grains}
\label{sec:magn-effects-grain}

It remains to calculate in detail the effects on grain dynamics of the
plasma's magnetic field, in order to justify the approach taken in
\S~\ref{sec:tight-magn-coupl} and extend those results to include the
effects of drag forces from the gas.  The Lorentz force on charged
grains due to a magnetic field is
\begin{equation}
  \label{eq:f-lorentz}
  \bm{f}\!\B = \frac{z\grain e}{c} \, \bm{w} \times \bm{B} \ . 
\end{equation}
The direction of the force is perpendicular both to the magnetic
field, \(\bm{B}\), and to the relative velocity, \(\bm{w}\), of the
grain with respect to the plasma.  If \(\bm{w}\) and \(\bm{B}\) (as
seen by the grain) are changing slowly, compared with the
gyrofrequency, \(\omega\B = z\grain e B / m\grain c\), then the grain
motion perpendicular to \(\bm{B}\) is constrained to be a circle of
radius equal to the Larmor radius:
\begin{equation}
  \label{eq:Larmor}
  r\B = \frac{m\grain c w_\perp} {\abs{z\grain} e B} \ ,
\end{equation}
where \(B = \abs{\bm{B}}\) and \(w_\perp\) is the perpendicular component
of \(\bm{w}\).  The component of \(\bm{w}\) parallel to \(\bm{B}\) is
unaffected by \(\bm{f}\!\B\), so the resultant trajectory is helical.

The relative importance of the magnetic field can be characterized by
the ratio of the Larmor radius to the minimum radius, \(\Rmin\),
reached by the grain in the dust wave (see
\S~\ref{sec:grain-traj-along}), where \(\Rmin \approx R_\dag\) for
drag-confined dust waves (DDW), or \(\Rmin \approx R\starstar\) for
inertia-confined dust waves (IDW).  We write the field strength in
terms of the Alfvén speed,
\begin{equation}
  \label{eq:alfven}
  v\alfven = \frac{B}{(4\pi\rho\gas)^{1/2}}
  = 1.9\, \frac{B}{\si{\micro G}} n^{-1/2} \, \si{km.s^{-1}} \ ,
\end{equation}
and the grain charge \(z\grain e\) in terms of the potential \(\phi\) (eq.~[\ref{eq:phi-potential}]) to obtain
\begin{equation}
  \label{eq:larmor-over-Rdag}
  \text{DDW:}\quad \frac{r\B}{\Rmin} =  
  \frac{r\B}{R_\dag} = 0.0140 \,
  a_{\si{\um}}^2 \,
  \frac{w_\perp}{v\alfven} \,
  \left( \frac{\Xi_\dag}{L_4 T_4} \right)^{1/2}
  \frac{\rho\grain}{\phi_\dag}
\end{equation}
and
\begin{equation}
  \label{eq:larmor-over-Rstarstar}
  \text{IDW:}\quad \frac{r\B}{\Rmin} =  
  \frac{r\B}{R\starstar} = 0.0544 \,
  a_{\si{\um}}^3 \,
  \frac{w_\perp}{v\alfven} \,
  \frac{v_{10}^2 }{n^{1/2}} \,
  \frac{1}{L_4 T_4} \,
  \frac{\rho\grain^2}{\Qp \phi\starstar} \ ,
\end{equation}
where \(a_{\si{\um}} = a / \SI{1}{\um}\), \(\rho\grain\) is the grain
material density in \si{g.cm^{-3}}, and we have made use of
equations~(\ref{eq:Rstar-typical}, \ref{eq:dust-r0}, \ref{eq:Rdag-over-Rstar}).

If \(r\B / \Rmin \ll 1\), then the grains are so strongly coupled to the
field that they can be treated in the guiding-center approximation, in
which the trajectory is decomposed into a tight circular gyromotion
around the field lines, plus a sliding of the guiding center along the
field lines, which is governed by the radiation and drag forces.  The
radiation force will also produce an out-of-plane drift, given by
\begin{equation}
  \label{eq:perpendicular-drift}
  \bm{v}_{\text{drift}} = \frac{c}{e z\grain} \, \frac{\bm{f}\!\rad \times \bm{B}}{B^2}
  \ ,
\end{equation}
but from equations~(\ref{eq:Larmor}, \ref{eq:dust-rad-force},
\ref{eq:dust-r0}) it follows that
\begin{equation}
  \label{eq:vdrift-over-vinfinity}
  \frac{v_{\text{drift}}(R\starstar)}{v_\infty} = \frac{r\B}{2 R\starstar} \ ,
\end{equation}
so it is valid to ignore this drift in the limit of small \(r\B\).
This is the limit that was applied in \S~\ref{sec:tight-magn-coupl}
for the case of zero drag.  In the opposite limit,
\(r\B / \Rmin \gg 1\), magnetic coupling is so weak that the
non-magnetic results of \S~\ref{sec:imperf-coupl-betw} are scarcely
modified.  Assuming \(w_\perp \sim v_\infty\) and adopting a threshold of
\(r\B / \Rmin < 0.1\), equations~(\ref{eq:larmor-over-Rdag},
\ref{eq:larmor-over-Rstarstar}) can be transformed into conditions on
the stream velocity (in \si{km.s^{-1}}) where tight magnetic coupling
will apply: \(v_{\infty} < v\freeze\), where
\begin{equation}
  \label{eq:velocity-strong-B-coupling}
  \begin{aligned}
    \text{drag-confined:}\quad v\freeze
    &\approx 0.8 \, a_{\si{\um}}^{-2} \, v\alfven \, L_4^{1/2} \ , \\
    \text{inertia-confined:}\quad v\freeze
    &\approx 6 \, a_{\si{\um}}^{-1} \, v\alfven^{1/3} \, n^{1/6} \, L_4^{1/3} \ ,
  \end{aligned}
\end{equation}
in which we have substituted typical values of the minor parameters
\(\Xi_\dag\), \(\phi_{\dag}\), \(\phi_{**}\), \(\rho\grain\),
\(T_4\).\footnote{%
  The most significant systematic variation in \(v\freeze\) from these
  suppressed parameters is due to grain composition, yielding slightly
  higher values for graphite than for silicate (\SI{+-0.15}{dex}).} %
It is apparent that \(v\freeze\) is very sensitive to the grain size.
For instance, taking a typical \hii{} region value of
\(v\alfven = \SI{2}{km.s^{-1}}\) \citep{Arthur:2011a} and
\(L_4 = 0.63\) (Tab.~\ref{tab:stars}, B1.5~V star), then for the
drag-confined case \(v\freeze \approx \SI{30}{km.s^{-1}}\) for \SI{0.2}{\um}
grains but \(v\freeze \approx \SI{3000}{km.s^{-1}}\) for \SI{0.02}{\um}
grains. Thus, for typical stream velocities of
\SIrange{20}{100}{km.s^{-1}}, the small grains are always tightly
coupled to the magnetic field, but the large grains are only loosely
coupled for the faster streams.

\subsection{Grain trajectories with tight magnetic coupling}
\label{sec:grain-traj-with}

\begin{figure*}
  \centering
  \includegraphics[width=\linewidth]{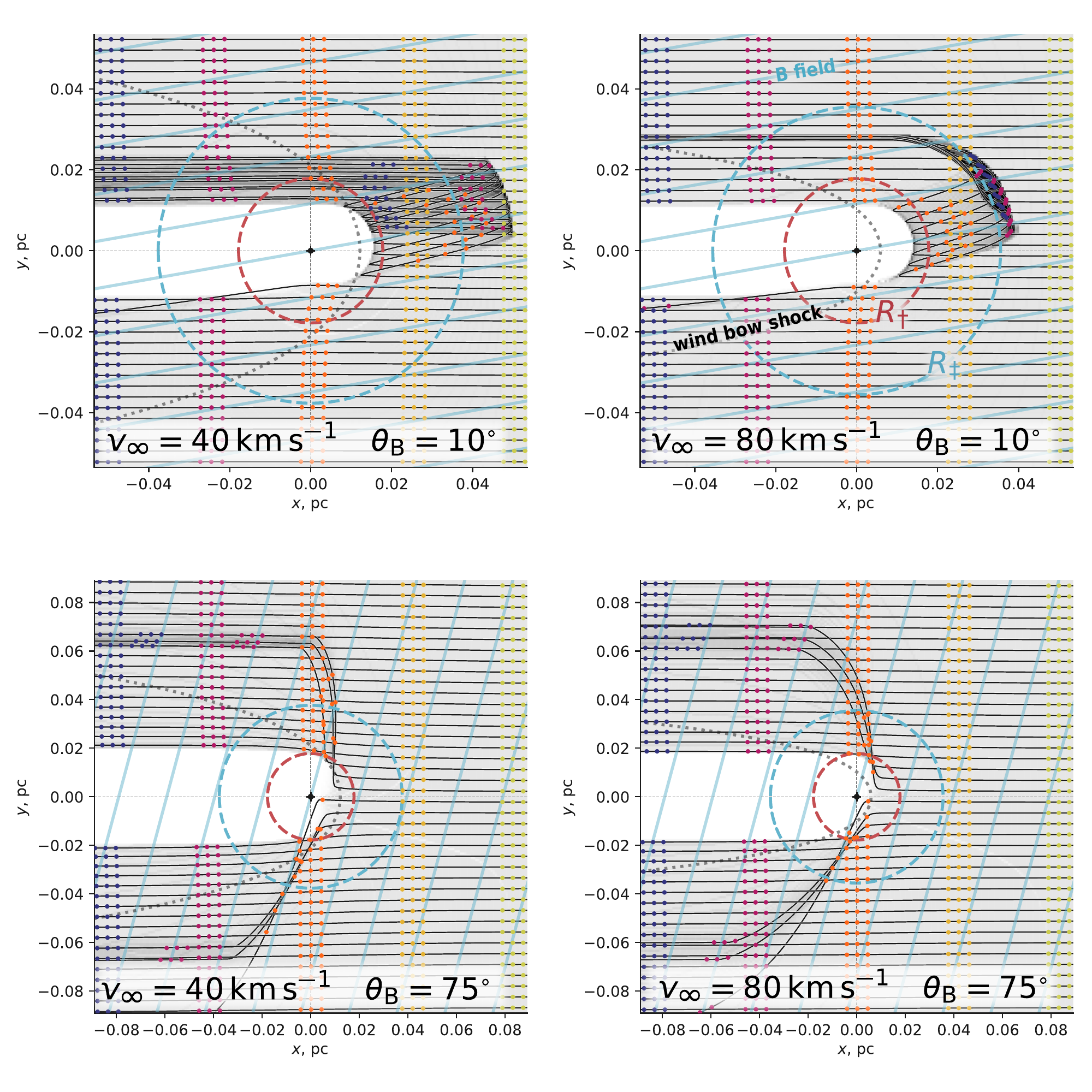}
  \caption{Drag-confined dust waves with tight magnetic coupling.
    Upper panels show a quasi-parallel field, \(\theta\B = \ang{10}\),
    while lower panels show a quasi-perpendicular field,
    \(\theta\B = \ang{75}\).  Left panels show an incident stream velocity of
    \(v_\infty = \SI{40}{km.s^{-1}}\), while right panels show
    \(v_\infty = \SI{80}{km.s^{-1}}\).  In all cases, the stream density is
    \(n = \SI{10}{cm^{-3}}\) and the calculations are performed for
    small graphite grains, \(a = \SI{0.02}{\um}\), and the
    \SI{10}{M_\odot} main-sequence B~star.  Continuous black lines show
    grain trajectories, with triplets of colored symbols indicating
    the progress of individual cohorts, which entered from the right
    edge at a particular time.  Continuous blue lines show the
    magnetic field, which flows from right to left along with the
    incident stream.  The radius of the rip point, \(R_\dag\), and the
    stagnant drift point, \(R_\ddag\), are shown respectively by red and
    blue dashed lines.  The approximate shape of the wind-supported
    bow shock is shown by the dotted gray line.  The calculations are
    no longer valid after trajectories cross this surface.}
  \label{fig:frozen-stream}
\end{figure*}

\begin{figure}
  \centering
  \includegraphics[width=\linewidth]{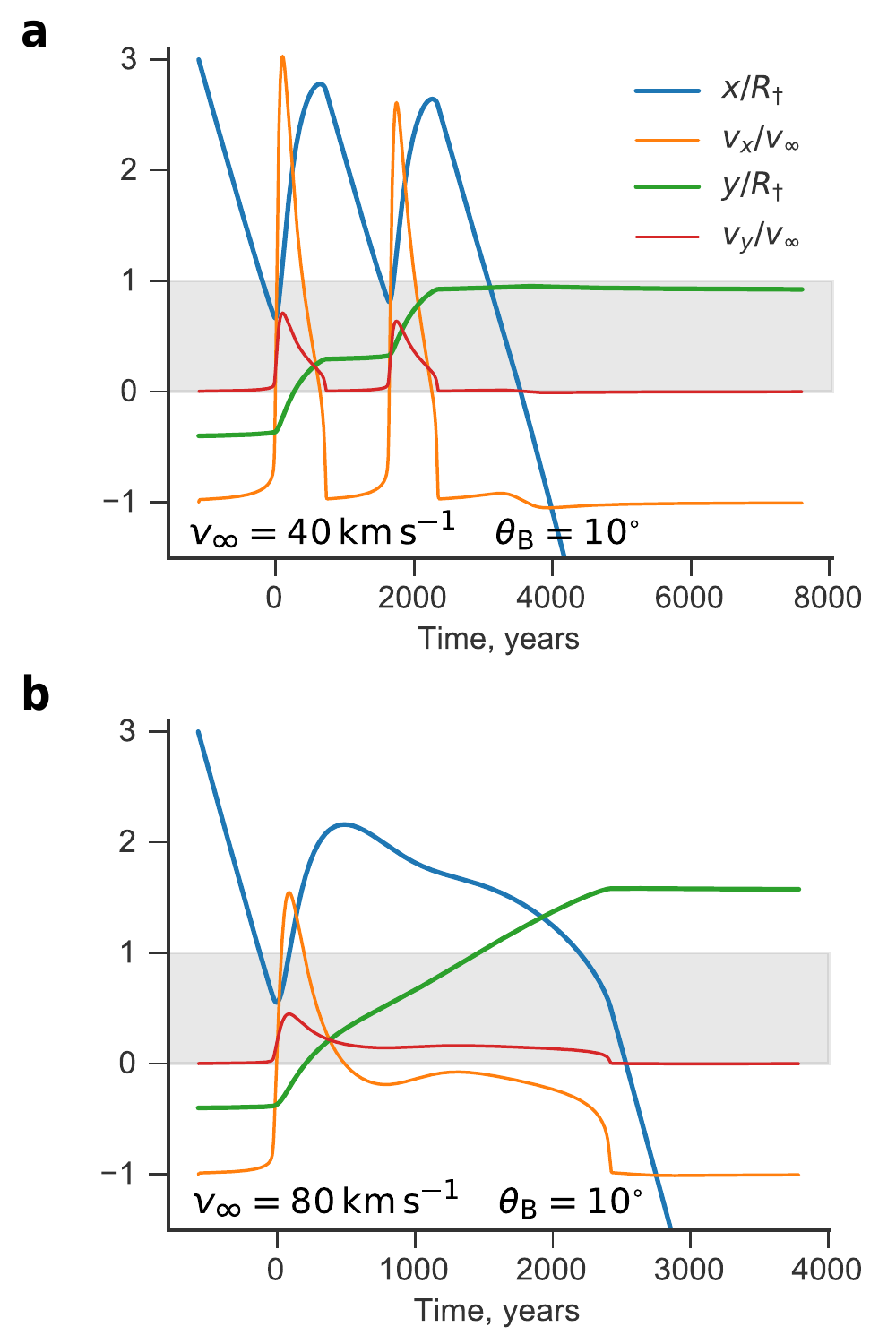}
  \caption{Sample grain trajectories for drag-confined dust waves with
    tight magnetic coupling and a quasi-parallel field
    orientation. These are the same models as in the upper row of
    Fig.~\ref{fig:frozen-stream}. (a)~Incident stream velocity of
    \(v_\infty = \SI{40}{km.s^{-1}}\), showing quasi-limit-cycle behavior
    (\numrange{0}{4000} years).  (b)~Incident stream velocity of
    \(v_\infty = \SI{80}{km.s^{-1}}\), showing quasi-stagnation behavior
    (\numrange{500}{2000} years).  In both cases, the streamline with
    initial impact parameter \(y = -0.4 R_\dag\) is shown.}
  \label{fig:frozen-trajectories}
\end{figure}

We can now investigate how the results of the
section~\ref{sec:imperf-coupl-betw} are modified by magnetic fields in
the tight coupling limit.  For simplicity, we assume a uniform field
in the incoming stream, with field lines oriented at an angle
\(\theta\B\) to the velocity vector that defines the bow axis.  We also
assume a super-alfvénic stream, \(v_\infty > v\alfven\), so that the
radius, \(R_0\), of the wind bow shock is unaffected by the magnetic
field, and additionally assume \(\tau_* \ll \tau_{*, \text{max}}\), so that
the back-reaction of the grain drag on the plasma is negligible
(\S~\ref{sec:back-reaction-gas}) and \(B\) remains uniform in
magnitude and direction in the dust wave region, outside of the bow
shock.

In \S~\ref{sec:tight-magn-coupl}, we derived analytic and
semi-analytic results in the limit of zero gas--grain drag, which is
appropriate for inertia-confined dust waves.  The resultant dust wave
structure is highly dependent on the field orientation.  For a
parallel field (Fig.~\ref{fig:inertia-thB0}), the apex of the dust
wave occurs at the same point, \(R\starstar\), as in the non-magnetic
case, but the shape of the dust wave wings is more closed, being
hemispherical rather than parabolic in shape.  For a perpendicular
field (Fig.~\ref{fig:inertia-thB90}), on the other hand, grains in the
apex region are dragged very close to the star and no dust wave forms
there.  A dust wave can form in the wings, with impact parameter
\(> R\starstar\), which is roughly parabolic in shape, but more
swept-back than in the non-magnetic case.  Whether such a dust wave
will exist in practice depends critically on the size and shape of the
interior MHD wind-supported bow shock.

In Figures~\ref{fig:frozen-stream} and~\ref{fig:frozen-trajectories}
we show example results for drag-confined dust waves, which are
calculated by numerically integrating the grain's equation of motion,
as described in Appendix~\ref{sec:equat-moti-grains}. Apart from the
inclusion of the magnetic field, the model parameters are the same as
used in Figure~\ref{fig:phase-space-trajectories}, with the exception
that the stream density is increased to \SI{10}{cm^{-3}}.\footnote{The
  reason for using a higher density is to decrease the amplitude of
  the radial oscillations of the trajectories, which allows the dust
  wave structure to be more clearly perceived in the figures. } This
time, we use quasi-parallel (\(\theta\B = \ang{10}\)) and
quasi-perpendicular (\(\theta\B = \ang{75}\)) field orientations.  The
quasi-parallel field is most similar to the non-magnetic case, and the
models shown in the two upper panels of Figure~\ref{fig:frozen-stream}
closely mirror the cases shown in
Figure~\ref{fig:phase-space-trajectories}, with a limit cycle behavior
when \(v_\infty = \SI{40}{km.s^{-1}}\) and stagnant drift when
\(v_\infty = \SI{80}{km.s^{-1}}\).

The principle difference from the 1-D axial trajectories discussed in
\S~\ref{sec:grain-traj-along} is that the small \ang{10} misalignment
of \(\bm{B}\) from the incident stream direction causes a slow
sideways migration, which puts a finite limit on the time a grain can
reside in front of the star.  This can be appreciated more clearly in
Figure~\ref{fig:frozen-trajectories}, which shows the grain position
and velocity along a sample streamline with initial impact parameter
\(b = -0.4 R_\dag\) for the two quasi-parallel models.  In panel~a,
corresponding to \(v_\infty = \SI{40}{km.s^{-1}}\), we see the same
rip-and-snap-back cycle of de-coupling and re-coupling that was
discussed previously, but only two periods of the cycle are completed
before the grain's lateral migration takes it as far as
\(y \approx +R_\dag\), at which point nothing can stop the gas stream from
dragging it past the star and away.  All told, the grain remains in
the apex region for about \(1/\sin\theta\B\) times longer than the crossing
time, \(R_\dag / v_\infty\).

In panel~b, corresponding to \(v_\infty = \SI{80}{km.s^{-1}}\), the grain
settles for a while around the stagnant drift radius, \(R_\ddag\), after
the initial rip and turn around.  Again, it slowly migrates sideways,
and eventually recouples to the incident stream, but this time after
reaching \(y \approx +R_\ddag\).  The grain residence time in the apex region,
measured in crossing times, is slightly longer than in panel~a, but is
of the same order.  In both these cases a second, exterior dust
shell is formed in addition to the hemispherical one produced by the
initial turn around inside \(R_\dag\). In the limit-cycle case, it forms
at the snap-back point, while in the stagnant-drift case, it forms at
\(R_\ddag\) and is significantly denser than the interior shell (upper
right panel of Fig.~\ref{fig:frozen-stream}).

The assumptions behind these models break down when the grain
trajectory intersects the outer shock of the dust-free wind-supported
bow shock.  The increased gas pressure in the bow shock shell will
reduce \(\Xi\), which is likely to cause gas--grain recoupling in the
case of drag-confined dust waves (see also discussion in the final
paragraph of \S~\ref{sec:exist-cond-separ}).  However, detailed
modeling of this requires magnetohydrodynamical simulations, which are
beyond the scope of this paper.  In Figure~\ref{fig:frozen-stream} we
show in gray dotted lines the Wilkinoid surface \citep{Wilkin:1996a},
which is an approximation to the shape of the inner bow shock, see
\citet{Tarango-Yong:2018a}.  It can be seen that the majority of dust
streamlines do not cross this surface until well into the wings of the
bow, so that a separate exterior dust wave can exist for this
quasi-parallel magnetic field orientation.

This is no longer the case for the quasi-perpendicular field
orientation, as illustrated in the lower panels of
Figure~\ref{fig:frozen-stream}, where the behavior is similar to the
perpendicular inertia-confined case studied in
\S~\ref{sec:perp-magn-field}.  Although the grains decouple from the
gas inside the rip point, for small impact parameters the magnetic
geometry does not allow the radiation field to expel them until they
are much nearer to the star.  Since the inner wind-supported bow shock
radius is only a few times smaller than \(R_\dag\), it is possible that
they will pass through the shock surface and re-couple before
expulsion can occur.  For the \SI{40}{km.s^{-1}} model (lower left
panel), all of the streamlines with impact parameters
\(\abs{b} < R_\dag\) intersect the Wilkinoid surface before they are
deflected, and so no separate dust wave would form in this case.  For
the \SI{80}{km.s^{-1}} model (lower left panel), some of the
streamlines (mainly those with \(b > 0\)) do manage to avoid crossing
the bow shock surface, so it is possible that a dust wave may still
form, although it would be on only one side of the axis.


%% file: sec-dust-wave-discussion.tex
\section{Discussion}
\label{sec:discussion}

In this section we briefly discuss our predictions for the
appearance of dust waves and place our results in the context of
previous work on gas--grain coupling in the environs of high-mass
stars.  We concentrate on conceptual and theoretical aspects,
postponing empirical questions about particular sources for later
papers.

\subsection{Predicted appearance of dust waves}
\label{sec:pred-shape-struct}
By definition, dust waves do not correspond to any structure in the
gas and so are only detectable via the mid-infrared thermal emission
of the grains.  The emissivity is proportional to the grain density
but also depends on the grain temperature, which is a decreasing
function of radius, \(R\), from the star \citetext{see, for example,
  synthetic observations in \citealp{Mackey:2016a, Acreman:2016a,
    Meyer:2017a}}.  In radiative equilibrium, the bolometric
emissivity is proportional to the stellar flux \(\propto R^{-2}\), but the
radial dependence of the monochromatic emissivity will be much steeper
than this on the short-wavelength side of the average grain thermal
spectrum.  With this in mind, we can crudely estimate the appearance
of the example dust waves calculated in \S\S~\ref{sec:gas-free-bow}
and \ref{sec:grain-traj-with}.

In the inertia-confined case, the radius \(R\starstar\) is
proportional to the single-grain opacity \(\kappa\grain\) and will
therefore vary with grain size and composition.  For a spherical grain
of size \(\amu\,\si{\um}\) and solid density
\(\rho\grain \, \si{g.cm^{-3}}\), equation~\eqref{eq:kappa-grain} yields
\begin{equation}
  \label{eq:kappa-d-dependence}
  \kappa\grain = 7500 \frac{\Qpbar}{\rho\grain \amu} \ . 
\end{equation}
The radiation pressure efficiency is \(\Qpbar \sim 1\) for
\(\amu > 0.02\), but falls as \(\Qpbar \propto a\) for smaller grains
\citep[e.g., Fig.~7 of][]{Draine:2011a}.  Therefore, all the smallest
grains will form the dust wave at the same point, but grains larger
than \SI{0.02}{\um} will be spread out with
\(R\starstar \propto a^{-1}\).  It might be thought that this would produce
a very broad diffuse appearance to the dust wave, but this is
mitigated by the fact that it is the smaller grains that dominate the
UV opacity and mid-IR emissivity.  Also, the larger grains may be
destroyed by radiative torques (see \S~\ref{sec:grain-survival-dust}).
For the drag-confined case, the situation is clearer since the dust
wave will form just inside the rip point, \(R_\dag\), which is relatively
insensitive to the grain size.

The effect of a roughly parallel magnetic field is mainly seen in the
shape of the dust wave, which becomes hemispherical
(Fig.~\ref{fig:inertia-thB0}) instead of parabolic as was found in the
non-magnetic case (Fig.~\ref{fig:dust-trajectories}).  Interestingly,
this effect is the opposite of what is seen in simulations of MHD bow
shocks \citep{Meyer:2016a}, where a parallel B-field leads to
flatter-nosed bow shapes with a high planitude \citetext{see Fig.~25
  of \citealp{Tarango-Yong:2018a}}.  Figure~\ref{fig:frozen-stream}ab
shows that the field orientation has a similar effect on the dust wave
shape for drag-confined magnetic dust waves, but with the added
complication that a second dense shell forms on one side of the axis
at the stagnant drift radius, \(R_\ddag\), in cases where
\(v_\infty > \SI{60}{km.s^{-1}}\).  However, since \(R_\ddag\) is several times
larger than \(R_\dag\), the emission from this outer shell is likely to
be weak, given the steep radial dependence of the mid-IR emissivity
discussed above.

When the magnetic field is close to perpendicular to the axis, then a
dense dust shell does not form in the apex region for either the
inertia-confined (Fig.~\ref{fig:inertia-thB90}) or drag-confined
(Fig.~\ref{fig:frozen-stream}cd) cases.  Nonetheless, it is possible
that a dust wave might form in the wings in such cases.

\subsection{Gas--grain dynamics in \hii{} regions}
\label{sec:gas-grain-dynamics-hii}

The distinction that we draw in \S~\ref{sec:exist-cond-separ} between
inertia-confined dust waves and drag-confined dust waves is novel.
However, the concept of the \textit{rip point} can be found in
previous works, although not under that name and it has generally been
assumed to be of little interest.  For instance, Figure~8 of
\citet{Draine:2011a} clearly shows the discontinuity in drift velocity
that occurs at small radii within an \hii{} region (differences from
our own Figs.~\ref{fig:multi-dustprops} and~\ref{fig:drift-gn} are due
to \citeauthor{Draine:2011a}'s assumption of constant grain
potential).  However, the region of supersonic drift involves a very
small fraction of the total dust in the \hii{} region, so it was not
relevant to the concerns of that paper. Similarly, in \S~3.1 of
\citet{Hopkins:2018c} the possibility is raised of a ``decoupling
instability'' in the case that coupling strength decreases with
increasing slip velocity, but they go on to dismiss this as physically
irrelevant.  On the contrary, we show in this paper that it \emph{is}
relevant so long as the grain potential is high, so that a distinct
local maximum occurs in the drag-versus-velocity profile (see
Fig.~\ref{fig:gas-grain-drag-photoionized}).  In \S~9 of
\citet{Hopkins:2018a}, various issues relating to grain charging and
drag are discussed.  The quantity
\(e_{\text{rad}} / e_{\text{therm}}\) from \S~9.1.4 of that paper is
the same as our radiation parameter, \(\Xi\).  In \S~9.2.3 they give an
expression for the critical radius that divides subsonic from
supersonic drift, which is exactly equivalent to our
equation~\eqref{eq:Rdag-over-Rstar} for the rip point radius.  The
parameters assumed by \citeauthor{Hopkins:2018a} yield
\(\Xi_\dag = 3200\), which is within the range of values that we find for
O~stars in Table~\ref{tab:Xi-rip}.

\citet{Akimkin:2017a} studied the effect of radiation pressure on the
time-dependent evolution of an \hii{} region, taking full account of
dynamical coupling between grains and gas, thus extending a previous
study \citep{Akimkin:2015a} that had ignored the back-reaction on the
gas (see our \S~\ref{sec:back-reaction-gas}).  They found that the
dust completely decouples from the gas in the zone around the star
(\(< \SI{0.2}{pc}\) on timescales \(\sim \SI{0.5}{Myr}\)), which produces
an inner dust hole but partially eliminates the inner gas density hole
that is found in models that assume perfect coupling
\citep{Mathews:1967a, Draine:2011a, Kim:2016b}. The decoupling is more
pronounced for stars of lower effective temperature, which is probably
related to the fact that we find the critical radiation parameter
\(\Xi_\dag\) to be lower for B~stars than for O~stars
(Table~\ref{tab:Xi-rip}).  A similar dust hole for the central region
is found by \citet{Ishiki:2018a}.

The gas--grain decoupling that we have discussed so far occurs at high
radiation parameter, \(\Xi \sim 1000\), but there is another regime of
potential decoupling that occurs at low \(\Xi\).  It has long been
recognized \citep{Gail:1979a} that Coulomb gas--grain coupling should
weaken in the outer zones of an \hii{} region due to the fact that the
grain potential passes through (or close to) zero as a result of
electron collisions becoming competitive with photoelectric ejection
as the radiation field weakens.  This is the regime around
\(\Xi = \text{\numrange{0.1}{1.0}}\) in Figure~\ref{fig:drift-gn},
giving drift velocities of order \SI{1}{km.s^{-1}} for larger grains
(\SI{0.2}{\um}), which may be sufficient to produce significant
spatial variations in grain abundance on Myr timescales
\citep{Ishiki:2018a}. It remains to be seen whether this is still true
once effects that have been neglected in existing models are accounted
for, such as the \hii{} region's magnetic field \citep{Krumholz:2007a,
  Arthur:2011a, Gendelev:2012a} and internal turbulence
\citep{Arthur:2016a}.  On the other hand, grain abundance variations
may be enhanced by the resonant drag instability \citep{Squire:2018a,
  Hopkins:2018a}.  Yet another regime of potential decoupling occurs
in the neutral shell outside the \hii{} region
\citep{Gustafsson:2018a}, but since the drift velocity in both these
cases is much less than interesting values of the stream velocity,
\(v_\infty\), neither low-\(\Xi\) regime is relevant to dust waves.

\subsection{Gas--grain dynamics in stellar bows}
\label{sec:gas-grain-dynamics-bs}

The term \textit{dust wave} was first coined to described the
radiative decoupling of grains from the plasma flowing past a massive
star by \citet{Ochsendorf:2014b}, who attempted to explain the
infrared emission arc around \(\sigma\)~Ori.  The concept was subsequently
applied to additional dust arcs in RCW~82 and RCW~120
\citep{Ochsendorf:2014a} and a refined model was applied to further
observations of \(\sigma\)~Ori \citep{Ochsendorf:2015a}.  These works
provided the inspiration for the present paper, where we have
attempted to take a more a priori and systematic approach to the
problem, including factors such as the Lorentz force that were
neglected by \citeauthor{Ochsendorf:2014b}.

The most detailed simulations to date of the combined dynamics of
dust, gas, and magnetic field in an OB bow was carried out by
\citet{Katushkina:2017a}, who followed the trajectories of dust
particles after passing through an MHD bow shock, under the influence
of the stellar radiation force and Lorentz force.  The gas is
decelerated in the shock, but the dust initially carries on with its
pre-shock motion, which produces a relative velocity with respect to
the plasma. The component of this velocity perpendicular to the
magnetic field induces gyration about the field lines
(cf~\S~\ref{sec:magn-effects-grain}), which forms filaments of
enhanced dust density that are oriented perpendicular to the bow axis
and with characteristic separation equal to the bulk plasma velocity
times the gyration period.  In \citet{Katushkina:2018a}, similar such
simulations are applied to observations of the runaway B1 supergiant
\(\kappa\)~Cas, moving at \(\approx \SI{30}{km.s^{-1}}\), in order to explain the
filaments of infrared dust emission seen at \SI{24}{\um}.  It is found
that the simulations can only fit the observations with very large
dust grains (\(\approx \SI{1}{\um}\)) and a very strong perpendicular
magnetic field \(v\alfven \approx \SI{20}{km.s^{-1}}\).  The stellar
parameters of \(\kappa\)~Cas are roughly those given in the last row of
Table~\ref{tab:stars} and the observed radius of the dust arc (assumed
to correspond to the astropause) is \SI{0.75}{pc} for a distance of
\SI{1}{kpc}.  The \citeauthor{Katushkina:2018a} simulations do not
explicitly include the Coulomb drag force on the dust grains, although
our own results (\S~\ref{sec:cloudy-models-dust}) suggest that this
must be important around \(\kappa\)~Cas.  \citet{Katushkina:2018a} argue
that small grains are swept out by stellar radiation before reaching
the bow shock, and this would indeed be true if drag forces were
negligible.  However, we find \(\Xi \approx 20\) for the
\(\kappa\)~Cas shell, implying strong gas--grain coupling and that even
small grains cannot be repelled by the radiation force.  The root
cause of our difference with these authors is that they are assuming a
grain potential \(\phi \sim 1\), as seen in the ISM in the Solar
neighborhood, as opposed to \(\phi \sim 10\), which is more appropriate to
the environs of an OB star.  We also note that even when the Lorentz
force vastly exceeds all other forces, it does not necessarily follow
that the other forces are unimportant.  For example, in the
magnetized dust wave models that we present in
\S~\ref{sec:tight-magn-coupl} and \S~\ref{sec:grain-traj-with}, the
Lorentz force is infinitely stronger than other forces.  Nevertheless,
the radiation and drag forces are crucial in determining the structure
of the dust waves, which is possible because the component of the
Lorentz force projected along the field lines is zero.

\subsection{Grain survival}
\label{sec:grain-survival-dust}

The survival of dust grains of different sizes in close proximity to
OB stars is something that must be considered.  Potential destruction
mechanisms include thermal evaporation, particle sputtering, and
radiative torque destruction.  Thermal evaporation requires that the
grain radiative equilibrium temperature exceed the sublimation
temperature (\SIrange{1400}{1750}{K}, depending on composition), which
occurs for radiative fluxes about \SI{e9} times higher than the
interstellar radiation field.  For a drag-confined dust wave, we have
a fixed radiation parameter \(\Xi_\dag \sim 1000\), so this becomes a
threshold on the gas density, requiring \(n > \SI{3e5}{cm^{-3}}\).
Combining this with the maximum allowed density for a dust wave
(eq.~[\ref{eq:dust-wave-high-density-condition}]), we also obtain a
condition on the stream velocity:
\(v_\infty > 170 \kappa_{600}^{0.5} L_4^{0.15} \, \si{km.s^{-1}}\).  Therefore,
thermal evaporation is generally unimportant in dust waves, except for
fast-moving stars in very high density environments.

Ion sputtering is only effective for collider kinetic energies in
excess of \SI{100}{eV} \citep{Draine:1995a, Field:1997a}, which is
significantly larger than thermal energies in photoionized gas.  It
therefore requires supersonic gas--grain slip velocities
\(w > \SI{75}{km.s^{-1}}\), but this does not necessarily imply that
the stream velocity need be quite so high.  For inertia-confined dust
waves, \(w\) has a maximum value of \(2 v_\infty\) and for drag-confined
dust waves it can be even higher (for instance, reaching
\(w \approx 4 v_\infty\) in Fig.~\ref{fig:frozen-trajectories}a), so that
\(v_\infty > \SI{30}{km.s^{-1}}\) is probably sufficient.  However, in
order to be destroyed by sputtering it is also necessary that a dust
grain of radius \(\amu\)\,\si{\um} should traverse a gas column
density of \(\approx \num{2e21}\, \amu \,\si{cm^{-2}}\)
\citep{Draine:2011a}.  For magnetic field orientations close to
parallel (which is the case that most favors dust wave formation), the
grains linger in the dust wave for several dynamical times (see
\S~\ref{sec:grain-traj-with}), so we estimate a total column of
\(\approx 10 n R_\dag\).  Using equations~(\ref{eq:Rstar-typical},
\ref{eq:Rdag-over-Rstar}) then implies that grains smaller than
\(\amu \sim 0.001 (L_4 n)^{1/2}\) can be destroyed by sputtering in the
dust wave.

Radiative torque disruption \citep{Hoang:2018a} is the centrifugal
rupture of an irregular grain of non-zero helicity that has been
suprathermally spun up by the anisotropic absorption of radiation
\citep{Dolginov:1976a, Draine:1996b, Lazarian:2007a}.  The process has
a very steep size dependence, being most effective in destroying
larger grains.  Using equation~(27) of \citet{Hoang:2018a} and
assuming a grain tensile strength of
\(S_{\text{max}} = \SI{2e10}{dyne.cm^{-2}}\) \citetext{\S~2.4 of
  \citealp{Borkowski:1995a}}, we find that grains larger than
\(\amu \sim 0.04\, n^{-0.123}\) are efficiently destroyed by this
mechanism at a distance \(R_\dag\) from the star. The spin-up timescales
are less than \(100\, n^{-1}\,\si{yr}\), which is short compared with
the dust wave dynamic timescale.

In summary, refractory grains in the size range
\SIrange{0.001}{0.04}{\um} are predicted to survive in dust waves.
Larger grains than this are unlikely to be able to resist
disintegration by radiative torques and smaller grains may be
destroyed by sputtering.



\section{Summary}
\label{sec:summary}

We have extended our previous study of bows around OB stars (Paper~I)
in order to consider the weak coupling case, in which a
radiation-supported dust wave decouples from the gas to form an
infrared emission arc outside of any hydrodynamic bow shock.  Our
principle findings are as follows:
\begin{enumerate}[1.]
\item Dust waves can only exist when the star's relative velocity with
  respect to its environment exceeds a critical value
  \(v_\infty > v_{\text{min}}\)
  (eq.~[\ref{eq:dust-wave-velocity-condition}]).  For O~stars with
  strong winds,
  \(v_{\text{min}} = \text{\SIrange{150}{300}{km.s^{-1}}}\), although
  for weak-wind stars and B~stars it can be as low as
  \SI{30}{km.s^{-1}}.
\item Additionally, the ambient density is constrained to lie within a
  certain range, \(n_{\text{min}} \to n_{\text{max}}\). For the lowest
  allowed relative velocities, \(v_\infty \approx v_{\text{min}}\), these are
  \(n_{\text{min}} = \SI{0.01}{cm^{-3}}\),
  \(n_{\text{max}} = \SI{100}{cm^{-3}}\) for B stars and
  \(n_{\text{min}} = \SI{1}{cm^{-3}}\),
  \(n_{\text{max}} = \SI{e5}{cm^{-3}}\) for strong-wind O~stars (see
  Fig.~\ref{fig:existence-dust-wave}).  Both these limits increase for
  higher velocities, as \(n_{\text{min}} \propto v_\infty^2\) and
  \(n_{\text{max}} \propto v_\infty^4\).
\item Dust waves may either be \textit{inertia-confined} or
  \textit{drag-confined}, where the inertia-confined regime (in which
  gas drag is always negligible) corresponds to a relatively narrow
  range of densities above \(n_{\text{min}}\).
\item In drag-confined dust waves, the gas--grain decoupling occurs
  suddenly at a \textit{rip point}, where the Coulomb drag
  catastrophically breaks down. The rip point occurs at a particular
  value of the radiation-to-gas pressure ratio:
  \(\Xi_\dag \sim 1000\), with little dependence on other parameters.
\item The post-rip grain trajectories are unstable for
  \(v_\infty = \text{\SIrange{10}{50}{km.s^{-1}}}\), exhibiting limit-cycle
  decoupling/recoupling behavior of repeated rip followed by
  snap-back.  For higher velocities a quasi-stationary stagnant drift
  shell can form on the axis.
\item Grain coupling to magnetic fields can modify these results, but
  this depends critically on the angle \(\theta\B\) between the field and
  the star's relative velocity vector.  For the quasi-parallel case
  (\(\theta\B < \ang{30}\)), the axial structure of the dust wave is
  little-changed, but the shape of the dust wave wings become more
  closed (hemispherical) than in the non-magnetic case.  For the
  quasi-perpendicular case (\(\theta\B > \ang{60}\)), a dust wave cannot form
  on the axis, although it is possible it may do so in the wings.
\end{enumerate}


%% file: app-dust-equations.tex
\section{Equation of motion for grains with radiation, gas drag, and
  magnetic field}
\label{sec:equat-moti-grains}

Following \citet{Draine:1979a}, the drag force on a grain that is
moving at relative velocity \(\bm{w} = \bm{v}\grain - \bm{v}\gas\)
through a partially ionized gas can be written as a sum over each
collider species, \(k\), with mass \(m_k\), abundance relative to
hydrogen \(\alpha_k\) and charge \(z_k\). If the relative speed,
\(w = \abs{\bm{w}}\), is normalized to the thermal speed of each
species:
\begin{equation}
  \label{eq:s-velocity}
  s_k = \left( m_k w^2 / 2 k T  \right)^{1/2} \ ,
\end{equation}
then the magnitude of the force is 
\begin{equation}
  \label{eq:ds79}
  f\drag = f_* \sum_k \alpha_k \left[ G_0(s_k) + z_k^2 \phi^2 \ln(\Lambda/z_k) G_2(s_k) \right],
\end{equation}
where \(f_*\) is a characteristic thermal force on the grain (see
eq.~[\ref{eq:fstar}]). The dimensionless functions of normalized speed
\(G_0(s)\) and \(G_2(s)\) are given by
\begin{align}
  \label{eq:G0}
  G_0(s) & = \left( s^2 + 1 - \frac{1}{4 s^2} \right) \erf(s)
  +  \left( s + \frac{1}{2 s} \right) \frac{e^{-s^2}}{\sqrt{\pi}}\\
  \label{eq:G2}
  G_2(s) & = \frac{\erf(s)}{s^2} - \frac{2 e^{-s^2}}{s \sqrt{\pi}} \ .
\end{align}
The \(G_0\) term is due to inelastic solid-body collisions in the
Epstein limit, and is derived in \S~4 of \citet{Baines:1965a}.  The
\(G_2\) term is due to electrostatic Coulomb interactions, with
\(\phi\) being the grain potential in thermal units
(eq.~[\ref{eq:phi-potential}]) and \(\Lambda\) the plasma parameter.  It was
first derived in the different context of dynamical friction in
stellar systems by \citet{Chandrasekhar:1941a}.

Figures~\ref{fig:drag-components}--\ref{fig:gas-grain-drag-photoionized}
show example applications of these equations to gas--grain drag in a
photoionized region.  The included collider species are protons,
electrons, helium ions, and metal ions.  Helium is assumed to be
singly ionized, leading to only a small contribution to the drag
force.  For much hotter stars, such as the central stars of planetary
nebulae, helium may be doubly ionized, which leads to a fourfold
increase in its Coulomb drag contribution, which is significant for
\(w < \SI{5}{km.s^{-1}}\).  All metals are lumped together as a single
species, assuming standard \hii{} region gas-phase abundances.  They
are dominated by C and O, with minor contributions from N and Ne.  The
total abundance is \num{8.5e-4} and the effective atomic weight is
\num{15.3}.  All are assumed to be doubly ionized.  Their largest
relative contribution to the drag force is for
\(w < \SI{2}{km.s^{-1}}\), but is less than 1\% even there.
  
The grain trajectories presented in \S~\ref{sec:grain-traj-along} and
\ref{sec:grain-traj-with} are calculated by numerically solving the
grain equation of motion:
\begin{equation}
  \label{eq:grain-equation-motion}
  m\grain \frac{d^2 \bm{r}}{d t^2} = \bm{f} \ .
\end{equation}
The total force \(\bm{f}\) is the sum of radiation, drag, and Lorentz
terms:
\begin{equation}
  \label{eq:total-force}
  \bm{f} = \frac{\sigma\grain \Qpbar L}{4\pi R^2 c} \hat{\bm{r}}
  - f\drag \hat{\bm{w}}
  + \frac{z\grain e}{c} \bm{w} \times \bm{B} \ ,
\end{equation}
with \(f\drag\) given by equation~\eqref{eq:ds79} and where
\(\hat{\bm{r}}\) is the unit vector in the radial direction and
\(\hat{\bm{w}} = \bm{w} / w\) is the unit vector along the direction
of gas--grain relative motion.  In the strong magnetic coupling limit
(see \S~\ref{sec:grain-traj-with} and \ref{sec:tight-magn-coupl}), the
Lorentz term is not included explicitly, but instead the equation of
motion is solved for the guiding center by replacing \(\bm{f}\) by its
projection along the magnetic field: 
\begin{equation}
  \label{eq:projected-force}
  \widetilde{\bm{f}} = (\bm{f} \cdot \hat{\bm{b}})  \, \hat{\bm{b}} \ ,
\end{equation}
where \(\hat{\bm{b}} = \bm{B} / B\).

If distances are measured in units of the radiative turnaround radius,
\(R\starstar\) (eq.~[\ref{eq:dust-r0}]), and times in units of
\(R\starstar / v_\infty\), then the grain acceleration
\(\bm{a}\grain = \bm{f} / m\grain\) in the non-magnetic case can be
written in non-dimensional form as
\begin{equation}
  \label{eq:grain-acceleration}
  \frac{\bm{a}\grain}{a\starstar}
  = \frac{R\starstar^2}{2 R^{2}} \hat{\bm{r}}
  - C\drag \frac{f\drag}{f_*} \hat{\bm{w}} \, 
\end{equation}
where \(a\starstar = v_\infty^2 / R\starstar\) is a characteristic
acceleration scale and the dimensionless drag constant is
\begin{equation}
  \label{eq:drag-constant}
  C\drag = \frac{4}{\Qpbar} \left(\frac{\sound \tau_* \kappa\grain}{v_\infty \kappa}\right)^2 \ .
\end{equation}

A collection of python programs that implement the equations of this
appendix is available at
\url{https://github.com/div-B-equals-0/dust-trajectories}, including
programs to generate all the grain trajectory figures of this paper
plus additional figures and movies.  The integration of
equation~\eqref{eq:grain-acceleration} is carried out using the
python library function \texttt{scipy.integrate.odeint}, which wraps
the Fortran ODEPACK library \citep{Hindmarsh:1983a, Jones:2001a}.

%% file: bs-bw-dw-02.bbl
\begin{thebibliography}{}
\makeatletter
\relax
\def\mn@urlcharsother{\let\do\@makeother \do\$\do\&\do\#\do\^\do\_\do\%\do\~}
\def\mn@doi{\begingroup\mn@urlcharsother \@ifnextchar [ {\mn@doi@}
  {\mn@doi@[]}}
\def\mn@doi@[#1]#2{\def\@tempa{#1}\ifx\@tempa\@empty \href
  {http://dx.doi.org/#2} {doi:#2}\else \href {http://dx.doi.org/#2} {#1}\fi
  \endgroup}
\def\mn@eprint#1#2{\mn@eprint@#1:#2::\@nil}
\def\mn@eprint@arXiv#1{\href {http://arxiv.org/abs/#1} {{\tt arXiv:#1}}}
\def\mn@eprint@dblp#1{\href {http://dblp.uni-trier.de/rec/bibtex/#1.xml}
  {dblp:#1}}
\def\mn@eprint@#1:#2:#3:#4\@nil{\def\@tempa {#1}\def\@tempb {#2}\def\@tempc
  {#3}\ifx \@tempc \@empty \let \@tempc \@tempb \let \@tempb \@tempa \fi \ifx
  \@tempb \@empty \def\@tempb {arXiv}\fi \@ifundefined
  {mn@eprint@\@tempb}{\@tempb:\@tempc}{\expandafter \expandafter \csname
  mn@eprint@\@tempb\endcsname \expandafter{\@tempc}}}

\bibitem[\protect\citeauthoryear{{Abel}, {Hoof}, {Shaw}, {Ferland}  \&
  {Elwert}}{{Abel} et~al.}{2008}]{Abel:2008a}
{Abel} N.~P.,  {Hoof} P.~A.~M.~v.,  {Shaw} G.,  {Ferland} G.~J.,   {Elwert} T.,
   2008, \apj, 686, 1125

\bibitem[\protect\citeauthoryear{{Acreman}, {Stevens}  \& {Harries}}{{Acreman}
  et~al.}{2016}]{Acreman:2016a}
{Acreman} D.~M.,  {Stevens} I.~R.,   {Harries} T.~J.,  2016, \mnras, 456, 136

\bibitem[\protect\citeauthoryear{{Akimkin}, {Kirsanova}, {Pavlyuchenkov}  \&
  {Wiebe}}{{Akimkin} et~al.}{2015}]{Akimkin:2015a}
{Akimkin} V.~V.,  {Kirsanova} M.~S.,  {Pavlyuchenkov} Y.~N.,   {Wiebe} D.~S.,
  2015, \mnras, 449, 440

\bibitem[\protect\citeauthoryear{{Akimkin}, {Kirsanova}, {Pavlyuchenkov}  \&
  {Wiebe}}{{Akimkin} et~al.}{2017}]{Akimkin:2017a}
{Akimkin} V.~V.,  {Kirsanova} M.~S.,  {Pavlyuchenkov} Y.~N.,   {Wiebe} D.~S.,
  2017, \mnras, 469, 630

\bibitem[\protect\citeauthoryear{{Ali}, {Sabin}, {Snaid}  \& {Basurah}}{{Ali}
  et~al.}{2012}]{Ali:2012a}
{Ali} A.,  {Sabin} L.,  {Snaid} S.,   {Basurah} H.~M.,  2012, \aap, 541, A98

\bibitem[\protect\citeauthoryear{{Arthur}, {Kurtz}, {Franco}  \&
  {Albarrán}}{{Arthur} et~al.}{2004}]{Arthur:2004a}
{Arthur} S.~J.,  {Kurtz} S.~E.,  {Franco} J.,   {Albarrán} M.~Y.,  2004, \apj,
  608, 282

\bibitem[\protect\citeauthoryear{{Arthur}, {Henney}, {Mellema}, {de Colle}  \&
  {V{\'a}zquez-Semadeni}}{{Arthur} et~al.}{2011}]{Arthur:2011a}
{Arthur} S.~J.,  {Henney} W.~J.,  {Mellema} G.,  {de Colle} F.,
  {V{\'a}zquez-Semadeni} E.,  2011, \mnras, 414, 1747

\bibitem[\protect\citeauthoryear{{Arthur}, {Medina}  \& {Henney}}{{Arthur}
  et~al.}{2016}]{Arthur:2016a}
{Arthur} S.~J.,  {Medina} S.-N.~X.,   {Henney} W.~J.,  2016, \mnras, 463, 2864

\bibitem[\protect\citeauthoryear{{Astropy Collaboration} et~al.,}{{Astropy
  Collaboration} et~al.}{2013}]{Astropy-Collaboration:2013a}
{Astropy Collaboration} et~al., 2013, \aap, 558, A33

\bibitem[\protect\citeauthoryear{{Astropy Collaboration} et~al.,}{{Astropy
  Collaboration} et~al.}{2018}]{Astropy-Collaboration:2018a}
{Astropy Collaboration} et~al., 2018, \aj, 156, 123

\bibitem[\protect\citeauthoryear{{Baines}, {Williams}  \& {Asebiomo}}{{Baines}
  et~al.}{1965}]{Baines:1965a}
{Baines} M.~J.,  {Williams} I.~P.,   {Asebiomo} A.~S.,  1965, \mnras, 130, 63

\bibitem[\protect\citeauthoryear{{Baldwin}, {Ferland}, {Martin}, {Corbin},
  {Cota}, {Peterson}  \& {Slettebak}}{{Baldwin} et~al.}{1991}]{Baldwin:1991a}
{Baldwin} J.~A.,  {Ferland} G.~J.,  {Martin} P.~G.,  {Corbin} M.~R.,  {Cota}
  S.~A.,  {Peterson} B.~M.,   {Slettebak} A.,  1991, \apj, 374, 580

\bibitem[\protect\citeauthoryear{{Birnstiel}, {Dullemond}  \&
  {Brauer}}{{Birnstiel} et~al.}{2010}]{Birnstiel:2010a}
{Birnstiel} T.,  {Dullemond} C.~P.,   {Brauer} F.,  2010, \aap, 513, A79

\bibitem[\protect\citeauthoryear{{Blaauw}}{{Blaauw}}{1961}]{Blaauw:1961a}
{Blaauw} A.,  1961, \bain, 15, 265

\bibitem[\protect\citeauthoryear{Bohren \& Huffman}{Bohren \&
  Huffman}{1983}]{Bohren:1983a}
Bohren C.~F.,  Huffman D.,  1983, Absorption and scattering of light by small
  particles.
Wiley-VCH

\bibitem[\protect\citeauthoryear{{Borkowski} \& {Dwek}}{{Borkowski} \&
  {Dwek}}{1995}]{Borkowski:1995a}
{Borkowski} K.~J.,  {Dwek} E.,  1995, \apj, 454, 254

\bibitem[\protect\citeauthoryear{{Brownsberger} \& {Romani}}{{Brownsberger} \&
  {Romani}}{2014}]{Brownsberger:2014a}
{Brownsberger} S.,  {Romani} R.~W.,  2014, \apj, 784, 154

\bibitem[\protect\citeauthoryear{{Cantó}, {Raga}  \& {González}}{{Cantó}
  et~al.}{2005}]{Canto:2005a}
{Cantó} J.,  {Raga} A.~C.,   {González} R.,  2005, \rmxaa, 41, 101

\bibitem[\protect\citeauthoryear{{Chandrasekhar}}{{Chandrasekhar}}{1941}]{Chandrasekhar:1941a}
{Chandrasekhar} I.~S.,  1941, \apj, 93, 285

\bibitem[\protect\citeauthoryear{{Contreras} \& {Rodr{\'{\i}}guez}}{{Contreras}
  \& {Rodr{\'{\i}}guez}}{1999}]{Contreras:1999a}
{Contreras} M.~E.,  {Rodr{\'{\i}}guez} L.~F.,  1999, \apj, 515, 762

\bibitem[\protect\citeauthoryear{{Cordes}, {Romani}  \& {Lundgren}}{{Cordes}
  et~al.}{1993}]{Cordes:1993a}
{Cordes} J.~M.,  {Romani} R.~W.,   {Lundgren} S.~C.,  1993, \nat, 362, 133

\bibitem[\protect\citeauthoryear{{Cox} et~al.,}{{Cox} et~al.}{2012}]{Cox:2012a}
{Cox} N.~L.~J.,  et~al., 2012, \aap, 537, A35

\bibitem[\protect\citeauthoryear{{Dipierro}, {Laibe}, {Alexander}  \&
  {Hutchison}}{{Dipierro} et~al.}{2018}]{Dipierro:2018a}
{Dipierro} G.,  {Laibe} G.,  {Alexander} R.,   {Hutchison} M.,  2018, \mnras,
  479, 4187

\bibitem[\protect\citeauthoryear{{Dolginov} \& {Mitrofanov}}{{Dolginov} \&
  {Mitrofanov}}{1976}]{Dolginov:1976a}
{Dolginov} A.~Z.,  {Mitrofanov} I.~G.,  1976, \apss, 43, 291

\bibitem[\protect\citeauthoryear{{Draine}}{{Draine}}{1995}]{Draine:1995a}
{Draine} B.~T.,  1995, \apss, 233, 111

\bibitem[\protect\citeauthoryear{{Draine}}{{Draine}}{2011}]{Draine:2011a}
{Draine} B.~T.,  2011, \apj, 732, 100

\bibitem[\protect\citeauthoryear{{Draine} \& {Salpeter}}{{Draine} \&
  {Salpeter}}{1979}]{Draine:1979a}
{Draine} B.~T.,  {Salpeter} E.~E.,  1979, \apj, 231, 77

\bibitem[\protect\citeauthoryear{{Draine} \& {Weingartner}}{{Draine} \&
  {Weingartner}}{1996}]{Draine:1996b}
{Draine} B.~T.,  {Weingartner} J.~C.,  1996, \apj, 470, 551

\bibitem[\protect\citeauthoryear{{Dzib}, {Rodr{\'{\i}}guez}, {Loinard},
  {Mioduszewski}, {Ortiz-León}  \& {Araudo}}{{Dzib} et~al.}{2013}]{Dzib:2013a}
{Dzib} S.~A.,  {Rodr{\'{\i}}guez} L.~F.,  {Loinard} L.,  {Mioduszewski} A.~J.,
  {Ortiz-León} G.~N.,   {Araudo} A.~T.,  2013, \apj, 763, 139

\bibitem[\protect\citeauthoryear{{Ferland} et~al.,}{{Ferland}
  et~al.}{2013}]{Ferland:2013a}
{Ferland} G.~J.,  et~al., 2013, \rmxaa, 49, 137

\bibitem[\protect\citeauthoryear{{Ferland} et~al.,}{{Ferland}
  et~al.}{2017}]{Ferland:2017a}
{Ferland} G.~J.,  et~al., 2017, \rmxaa, 53, 385

\bibitem[\protect\citeauthoryear{{Field}, {May}, {Pineau des Forets}  \&
  {Flower}}{{Field} et~al.}{1997}]{Field:1997a}
{Field} D.,  {May} P.~W.,  {Pineau des Forets} G.,   {Flower} D.~R.,  1997,
  \mnras, 285, 839

\bibitem[\protect\citeauthoryear{{Gail} \& {Sedlmayr}}{{Gail} \&
  {Sedlmayr}}{1979}]{Gail:1979a}
{Gail} H.~P.,  {Sedlmayr} E.,  1979, \aap, 77, 165

\bibitem[\protect\citeauthoryear{{Garc{\'{\i}}a-Arredondo}, {Henney}  \&
  {Arthur}}{{Garc{\'{\i}}a-Arredondo} et~al.}{2001}]{Garcia-Arredondo:2001a}
{Garc{\'{\i}}a-Arredondo} F.,  {Henney} W.~J.,   {Arthur} S.~J.,  2001, \apj,
  561, 830

\bibitem[\protect\citeauthoryear{{Geballe}, {Rigaut}, {Roy}  \&
  {Draine}}{{Geballe} et~al.}{2004}]{Geballe:2004a}
{Geballe} T.~R.,  {Rigaut} F.,  {Roy} J.-R.,   {Draine} B.~T.,  2004, \apj,
  602, 770

\bibitem[\protect\citeauthoryear{{Gendelev} \& {Krumholz}}{{Gendelev} \&
  {Krumholz}}{2012}]{Gendelev:2012a}
{Gendelev} L.,  {Krumholz} M.~R.,  2012, \apj, 745, 158

\bibitem[\protect\citeauthoryear{{Gull} \& {Sofia}}{{Gull} \&
  {Sofia}}{1979}]{Gull:1979a}
{Gull} T.~R.,  {Sofia} S.,  1979, \apj, 230, 782

\bibitem[\protect\citeauthoryear{{Gustafsson}}{{Gustafsson}}{2018}]{Gustafsson:2018a}
{Gustafsson} B.,  2018, \aap, 616, A91

\bibitem[\protect\citeauthoryear{{Henney} \& {Arthur}}{{Henney} \&
  {Arthur}}{2019}]{Henney:2019a}
{Henney} W.~J.,  {Arthur} S.~J.,  2019, arXiv e-prints, \href
  {http://adsabs.harvard.edu/abs/2019arXiv190303737H} {1903.03737 MNRAS
  submitted (Paper I)}

\bibitem[\protect\citeauthoryear{{Henney}, {Garc{\'{\i}}a-D{\'{\i}}az},
  {O'Dell}  \& {Rubin}}{{Henney} et~al.}{2013}]{Henney:2013a}
{Henney} W.~J.,  {Garc{\'{\i}}a-D{\'{\i}}az} M.~T.,  {O'Dell} C.~R.,   {Rubin}
  R.~H.,  2013, \mnras, 428, 691

\bibitem[\protect\citeauthoryear{Hindmarsh}{Hindmarsh}{1983}]{Hindmarsh:1983a}
Hindmarsh A.~C.,  1983, IMACS Transactions on Scientific Computation, 1, 55

\bibitem[\protect\citeauthoryear{{Hoang}, {Tram}, {Lee}  \& {Ahn}}{{Hoang}
  et~al.}{2018}]{Hoang:2018a}
{Hoang} T.,  {Tram} L.~N.,  {Lee} H.,   {Ahn} S.-H.,  2018, arXiv, \href
  {http://adsabs.harvard.edu/abs/2018arXiv181005557H} {1810.05557}

\bibitem[\protect\citeauthoryear{{Hopkins} \& {Lee}}{{Hopkins} \&
  {Lee}}{2016}]{Hopkins:2016a}
{Hopkins} P.~F.,  {Lee} H.,  2016, \mnras, 456, 4174

\bibitem[\protect\citeauthoryear{{Hopkins} \& {Squire}}{{Hopkins} \&
  {Squire}}{2018a}]{Hopkins:2018a}
{Hopkins} P.~F.,  {Squire} J.,  2018a, \mnras, 479, 4681

\bibitem[\protect\citeauthoryear{{Hopkins} \& {Squire}}{{Hopkins} \&
  {Squire}}{2018b}]{Hopkins:2018c}
{Hopkins} P.~F.,  {Squire} J.,  2018b, \mnras, 480, 2813

\bibitem[\protect\citeauthoryear{{Ishiki}, {Okamoto}  \& {Inoue}}{{Ishiki}
  et~al.}{2018}]{Ishiki:2018a}
{Ishiki} S.,  {Okamoto} T.,   {Inoue} A.~K.,  2018, \mnras, 474, 1935

\bibitem[\protect\citeauthoryear{Jones, Oliphant, Peterson  et~al.}{Jones
  et~al.}{2019}]{Jones:2001a}
Jones E.,  Oliphant T.,  Peterson P.,   et~al., 2001--2019, {SciPy}: Open
  source scientific tools for {Python}, \url {http://www.scipy.org/}

\bibitem[\protect\citeauthoryear{{Katushkina}, {Alexashov}, {Izmodenov}  \&
  {Gvaramadze}}{{Katushkina} et~al.}{2017}]{Katushkina:2017a}
{Katushkina} O.~A.,  {Alexashov} D.~B.,  {Izmodenov} V.~V.,   {Gvaramadze}
  V.~V.,  2017, \mnras, 465, 1573

\bibitem[\protect\citeauthoryear{{Katushkina}, {Alexashov}, {Gvaramadze}  \&
  {Izmodenov}}{{Katushkina} et~al.}{2018}]{Katushkina:2018a}
{Katushkina} O.~A.,  {Alexashov} D.~B.,  {Gvaramadze} V.~V.,   {Izmodenov}
  V.~V.,  2018, \mnras, 473, 1576

\bibitem[\protect\citeauthoryear{{Kim}, {Kim}  \& {Ostriker}}{{Kim}
  et~al.}{2016}]{Kim:2016b}
{Kim} J.-G.,  {Kim} W.-T.,   {Ostriker} E.~C.,  2016, \apj, 819, 137

\bibitem[\protect\citeauthoryear{{Kobulnicky}, {Schurhammer}, {Baldwin},
  {Chick}, {Dixon}, {Lee}  \& {Povich}}{{Kobulnicky}
  et~al.}{2017}]{Kobulnicky:2017a}
{Kobulnicky} H.~A.,  {Schurhammer} D.~P.,  {Baldwin} D.~J.,  {Chick} W.~T.,
  {Dixon} D.~M.,  {Lee} D.,   {Povich} M.~S.,  2017, \aj, 154, 201

\bibitem[\protect\citeauthoryear{{Krumholz}, {Stone}  \& {Gardiner}}{{Krumholz}
  et~al.}{2007}]{Krumholz:2007a}
{Krumholz} M.~R.,  {Stone} J.~M.,   {Gardiner} T.~A.,  2007, \apj, 671, 518

\bibitem[\protect\citeauthoryear{Landau \& Lifshitz}{Landau \&
  Lifshitz}{1976}]{Landau:1976a}
Landau L.,  Lifshitz E.,  1976, Mechanics, 3rd edn.
 Course of Theoretical Physics S Vol. 1, Butterworth-Heinemann

\bibitem[\protect\citeauthoryear{{Lanz} \& {Hubeny}}{{Lanz} \&
  {Hubeny}}{2003}]{Lanz:2003a}
{Lanz} T.,  {Hubeny} I.,  2003, \apjs, 146, 417

\bibitem[\protect\citeauthoryear{{Lanz} \& {Hubeny}}{{Lanz} \&
  {Hubeny}}{2007}]{Lanz:2007a}
{Lanz} T.,  {Hubeny} I.,  2007, \apjs, 169, 83

\bibitem[\protect\citeauthoryear{{Lazarian} \& {Hoang}}{{Lazarian} \&
  {Hoang}}{2007}]{Lazarian:2007a}
{Lazarian} A.,  {Hoang} T.,  2007, \mnras, 378, 910

\bibitem[\protect\citeauthoryear{{Lee}, {Hopkins}  \& {Squire}}{{Lee}
  et~al.}{2017}]{Lee:2017a}
{Lee} H.,  {Hopkins} P.~F.,   {Squire} J.,  2017, \mnras, 469, 3532

\bibitem[\protect\citeauthoryear{{Mackey}, {Haworth}, {Gvaramadze}, {Mohamed},
  {Langer}  \& {Harries}}{{Mackey} et~al.}{2016}]{Mackey:2016a}
{Mackey} J.,  {Haworth} T.~J.,  {Gvaramadze} V.~V.,  {Mohamed} S.,  {Langer}
  N.,   {Harries} T.~J.,  2016, \aap, 586, A114

\bibitem[\protect\citeauthoryear{{Mathews}}{{Mathews}}{1967}]{Mathews:1967a}
{Mathews} W.~G.,  1967, \apj, 147, 965

\bibitem[\protect\citeauthoryear{{Mattsson}, {Bhatnagar}, {Gent}  \&
  {Villarroel}}{{Mattsson} et~al.}{2019}]{Mattsson:2019a}
{Mattsson} L.,  {Bhatnagar} A.,  {Gent} F.~A.,   {Villarroel} B.,  2019,
  \mnras, 483, 5623

\bibitem[\protect\citeauthoryear{{Meyer}, {van Marle}, {Kuiper}  \&
  {Kley}}{{Meyer} et~al.}{2016}]{Meyer:2016a}
{Meyer} D.~M.-A.,  {van Marle} A.-J.,  {Kuiper} R.,   {Kley} W.,  2016, \mnras,
  459, 1146

\bibitem[\protect\citeauthoryear{{Meyer}, {Mignone}, {Kuiper}, {Raga}  \&
  {Kley}}{{Meyer} et~al.}{2017}]{Meyer:2017a}
{Meyer} D.~M.-A.,  {Mignone} A.,  {Kuiper} R.,  {Raga} A.~C.,   {Kley} W.,
  2017, \mnras, 464, 3229

\bibitem[\protect\citeauthoryear{{Ochsendorf} \& {Tielens}}{{Ochsendorf} \&
  {Tielens}}{2015}]{Ochsendorf:2015a}
{Ochsendorf} B.~B.,  {Tielens} A.~G.~G.~M.,  2015, \aap, 576, A2

\bibitem[\protect\citeauthoryear{{Ochsendorf}, {Cox}, {Krijt}, {Salgado},
  {Bern{\'e}}, {Bernard}, {Kaper}  \& {Tielens}}{{Ochsendorf}
  et~al.}{2014a}]{Ochsendorf:2014b}
{Ochsendorf} B.~B.,  {Cox} N.~L.~J.,  {Krijt} S.,  {Salgado} F.,  {Bern{\'e}}
  O.,  {Bernard} J.~P.,  {Kaper} L.,   {Tielens} A.~G.~G.~M.,  2014a, \aap,
  563, A65

\bibitem[\protect\citeauthoryear{{Ochsendorf}, {Verdolini}, {Cox}, {Bern{\'e}},
  {Kaper}  \& {Tielens}}{{Ochsendorf} et~al.}{2014b}]{Ochsendorf:2014a}
{Ochsendorf} B.~B.,  {Verdolini} S.,  {Cox} N.~L.~J.,  {Bern{\'e}} O.,  {Kaper}
  L.,   {Tielens} A.~G.~G.~M.,  2014b, \aap, 566, A75

\bibitem[\protect\citeauthoryear{{Povich}, {Benjamin}, {Whitney}, {Babler},
  {Indebetouw}, {Meade}  \& {Churchwell}}{{Povich} et~al.}{2008}]{Povich:2008a}
{Povich} M.~S.,  {Benjamin} R.~A.,  {Whitney} B.~A.,  {Babler} B.~L.,
  {Indebetouw} R.,  {Meade} M.~R.,   {Churchwell} E.,  2008, \apj, 689, 242

\bibitem[\protect\citeauthoryear{{Sanchez-Bermudez}, {Schödel}, {Alberdi},
  {Muzić}, {Hummel}  \& {Pott}}{{Sanchez-Bermudez}
  et~al.}{2014}]{Sanchez-Bermudez:2014a}
{Sanchez-Bermudez} J.,  {Schödel} R.,  {Alberdi} A.,  {Muzić} K.,  {Hummel}
  C.~A.,   {Pott} J.-U.,  2014, \aap, 567, A21

\bibitem[\protect\citeauthoryear{{Smith}, {Bally}, {Shuping}, {Morris}  \&
  {Kassis}}{{Smith} et~al.}{2005}]{Smith:2005a}
{Smith} N.,  {Bally} J.,  {Shuping} R.~Y.,  {Morris} M.,   {Kassis} M.,  2005,
  \aj, 130, 1763

\bibitem[\protect\citeauthoryear{{Spitzer}}{{Spitzer}}{1978}]{Spitzer:1978a}
{Spitzer} L.,  1978, {Physical processes in the interstellar medium}.
New York: Wiley-Interscience

\bibitem[\protect\citeauthoryear{{Squire} \& {Hopkins}}{{Squire} \&
  {Hopkins}}{2018}]{Squire:2018a}
{Squire} J.,  {Hopkins} P.~F.,  2018, \apjl, 856, L15

\bibitem[\protect\citeauthoryear{{Stevens}, {Blondin}  \& {Pollock}}{{Stevens}
  et~al.}{1992}]{Stevens:1992a}
{Stevens} I.~R.,  {Blondin} J.~M.,   {Pollock} A.~M.~T.,  1992, \apj, 386, 265

\bibitem[\protect\citeauthoryear{{Tarango-Yong} \& {Henney}}{{Tarango-Yong} \&
  {Henney}}{2018}]{Tarango-Yong:2018a}
{Tarango-Yong} J.~A.,  {Henney} W.~J.,  2018, \mnras, 477, 2431

\bibitem[\protect\citeauthoryear{{Tricco}, {Price}  \& {Laibe}}{{Tricco}
  et~al.}{2017}]{Tricco:2017a}
{Tricco} T.~S.,  {Price} D.~J.,   {Laibe} G.,  2017, \mnras, 471, L52

\bibitem[\protect\citeauthoryear{{Weidenschilling}}{{Weidenschilling}}{1977}]{Weidenschilling:1977b}
{Weidenschilling} S.~J.,  1977, \mnras, 180, 57

\bibitem[\protect\citeauthoryear{{Weingartner} \& {Draine}}{{Weingartner} \&
  {Draine}}{2001a}]{Weingartner:2001b}
{Weingartner} J.~C.,  {Draine} B.~T.,  2001a, \apjs, 134, 263

\bibitem[\protect\citeauthoryear{{Weingartner} \& {Draine}}{{Weingartner} \&
  {Draine}}{2001b}]{Weingartner:2001a}
{Weingartner} J.~C.,  {Draine} B.~T.,  2001b, \apj, 548, 296

\bibitem[\protect\citeauthoryear{{Weingartner}, {Draine}  \&
  {Barr}}{{Weingartner} et~al.}{2006}]{Weingartner:2006a}
{Weingartner} J.~C.,  {Draine} B.~T.,   {Barr} D.~K.,  2006, \apj, 645, 1188

\bibitem[\protect\citeauthoryear{{Wilkin}}{{Wilkin}}{1996}]{Wilkin:1996a}
{Wilkin} F.~P.,  1996, \apjl, 459, L31

\bibitem[\protect\citeauthoryear{{van Hoof}, {Weingartner}, {Martin}, {Volk}
  \& {Ferland}}{{van Hoof} et~al.}{2004}]{van-Hoof:2004a}
{van Hoof} P.~A.~M.,  {Weingartner} J.~C.,  {Martin} P.~G.,  {Volk} K.,
  {Ferland} G.~J.,  2004, \mnras, 350, 1330

\bibitem[\protect\citeauthoryear{{van Marle}, {Meliani}, {Keppens}  \&
  {Decin}}{{van Marle} et~al.}{2011}]{van-Marle:2011a}
{van Marle} A.~J.,  {Meliani} Z.,  {Keppens} R.,   {Decin} L.,  2011, \apjl,
  734, L26

\makeatother
\end{thebibliography}
